\documentclass[fleqn,usenatbib]{mnras}
\usepackage{amsmath}
\usepackage{amssymb}
\usepackage{graphicx}	

\usepackage{txfonts}
\newlength{\abovecaptionskip}%
\usepackage{threeparttable}

\def\msun{{$M_{\odot}$}}
\def\muphi{\mu_{\Phi}}
\def\muphim{{$\mu_{\Phi}$}}
\def\tff{t_{\rm ff}}
\def\tffm{{$t_{\rm ff}$}}
\newcommand{\be}{\begin{equation}}
\newcommand{\ee}{\end{equation}}

\def\eg{{\it e.g.,}}
\def\ie{{\rm i.e.,}}
\def\alfven{{Alf{\'v}en}}
\def\alfvenic{{Alf{\'v}enic}}
\def\calm{{\cal M}}
\def\ma{ {\calm_{\rm A}} }
\def\deg{{$^{\circ}$}}

\def\alphavir{\alpha_{\rm vir}}
\def\alphavirm{{$\alpha_{\rm vir}$}}
\def\bobs{{\beta_{\rm O}}}
\def\max{{\rm max}}
\def\mug{{$\mu$G}}
\def\mig{{mG}}

\def\muphieff{{\mu_{\Phi,\,\rm eff}}}
\def\nh{{n_{\rm H}}}
\newcommand{\ppbyp}[2]{{{\partial#1}\over{\partial#2}}}
\def\rcf{{r_{\rm cf}}}
\def\tot{{\rm tot}}

\title[Protostellar Disk Formation Aligned Turbulence]{Effect of Angular Momentum Alignment and Strong Magnetic Fields on the Formation of Protostellar Disks}

\author[William J. Gray et al.]{
William J. Gray,$^{1}$\thanks{E-mail: graywill@umich.edu}
Christopher F. McKee,$^{2,3}$
and Richard I. Klein,$^{3,4}$
\\
$^{1}$CLASP, College of Engineering, University of Michigan, 2455 Hayward St., Ann Arbor, Michigan 48109, USA\\
$^{2}$Department of Physics, University of California, Berkeley, Berkeley, CA 94720 \\
$^{3}$Department of Astronomy, University of California, Berkeley, Berkeley, CA 94720 \\
$^{4}$Lawrence Livermore National Laboratory, P.O. Box 808, L-023, Livermore, CA 94550
}
\date{Accepted 2017 September 13 . Received 2017 September 13 ; in original form 2017 June 29}
\pubyear{2017}
\begin{document}
\label{firstpage}
\pagerange{\pageref{firstpage}--\pageref{lastpage}}
\maketitle


\begin{abstract}
Star forming molecular clouds are observed to be both highly magnetized and turbulent. Consequently the formation of protostellar disks is largely dependent on the complex interaction between gravity, magnetic fields, and turbulence. Studies of non-turbulent protostellar disk formation with realistic magnetic fields have shown that these fields are efficient in removing angular momentum from the forming disks, preventing their formation. However, once turbulence is included, disks can form in even highly magnetized clouds, although the precise mechanism remains uncertain. Here we present several high resolution simulations of turbulent,
realistically magnetized, high-mass molecular clouds with both aligned and random turbulence to study the role that turbulence, misalignment, and magnetic fields have on the formation of protostellar disks. We find that when the turbulence is artificially aligned so that the angular momentum is parallel to the initial uniform field, no rotationally supported disks are formed, regardless of the initial turbulent energy. We conclude that turbulence and the associated misalignment between the angular momentum and the magnetic field are crucial in the formation of protostellar disks in the presence of realistic magnetic fields.

\end{abstract}

\section{Introduction}

The formation of a rotationally supported gaseous disk around protostars is one of the most important phases during the formation and evolution of protostars and for the formation of planets. Indeed they are consistently found around low-mass Class II young stellar objects (for a review see \cite{Williams2011}), and increasingly around Class I \citep[\eg][]{Jorgensen2009,Yen2013,Lee2011,Takakuwa2012}, and some Class 0 protostars \citep{Maury2010,Hara2013,Tobin2012,Tobin2013,Tobin2015,Tobin2016}. Class II sources in the Orion star forming region, for example, have protostellar disks with an average radius of 75 au and masses of 10$^{-2}$ \msun \citep{Williams2011}.

\citet{Tobin2015} state that there were four Class 0 sources with Keplerian disks known prior to their work. In their study of 9 Class 0 protostars in Perseus, they found three strong disk candidates with radii exceeding 100 au; however, their resolution was not adequate to determine if the disks were Keplerian. The median mass of the compact components (presumed to be disks) in their sample is $0.05$ M$_\odot$, larger than that inferred for disks associated with Class I and II sources. Therefore, the formation of a rotationally supported disk seems to be an ubiquitous feature of star formation.

It was originally thought that the formation of protostellar disks was a natural consequence of the conservation of angular momentum as a rotating molecular cloud undergoes gravitational collapse \citep[\eg][]{Bodenheimer1995,Lin1996}. If angular momentum is conserved during the collapse, an element of gas initially at radius $r_i$ rotating with angular velocity $\Omega$ will be in Keplerian rotation at the centrifugal radius (e.g., \citealt{Dapp2012})
\be
\rcf=\frac{j^2}{GM}=\frac{r_i^4\Omega^2}{GM},
\label{eq:rcf}
\ee
where $j$ is the specific angular momentum and $M$, the mass interior to the element of gas, is assumed to be spherically symmetric and to remain interior to the element of gas as the cloud contracts. It is common to characterize the importance of rotation by the parameter $\beta$, the ratio of rotational to gravitational energy. However, that cannot be directly observed. Instead, observers measure rotation by the value of $\beta$ for a uniform cloud,
\be
\bobs=\frac 13\left(\frac{R^3\Omega^2}{GM}\right),
\ee
where $M$ is the mass of the cloud; the value actually observed is $\bobs\sin^2 i$, where $i$ is the angle of inclination \citep{Goodman1993}. Real clouds are centrally concentrated, which increases the gravitational energy and decreases the rotational energy, so $\beta$ is generally less than $\bobs$ \citep{Goodman1993}.

Observed molecular clouds have a wide range of values of $\bobs$, with a typical value of 0.02 \citep{Goodman1993}. It follows that
\be
\frac{\rcf}{r_i}=3\bobs(r_i),
\ee
where $\bobs(r_i)$ is the value of $\bobs$ evaluated at $r_i$. For a typical value of $\bobs$, this means that a disk will form at a radius somewhat less than 10\% of the initial radius of the accreted gas, and indeed hydrodynamical collapse simulations without magnetic fields routinely find large rotation-dominated disks \citep{Machida2011}. However, observations of molecular clouds show that they are magnetized (for a review see \cite{Crutcher2012}). The addition of the magnetic field has a profound effect on the evolution and formation of protostellar disks.

The importance of the magnetic field is encapsulated in the mass-to-flux ratio relative to the critical value, \muphim\ \citep[][see Eq.~\ref{eqn:muphi} below]{Mouschovias1976}. \citet{Troland2008} estimate a median observed value $\muphi=1.3-2.6$ (we have corrected the lower bound to reflect the fact that projection effects can reduce the median $\muphi$ by up to a factor 4, not 3 as they assumed.) depending on whether the clumps in which the field is measured are disk-like or approximately spherical  In fact, the observed value of the magnetic field that enters this determination of \muphim\  is weighted by the density of the gas, whereas the value that enters \muphim\ is not; consequently, \citet{Li2015} infer that the actual median value of \muphim in molecular clumps is about 3.

As outlined by \cite{Joos2012}, \muphim\ measures the effectiveness of the magnetic field in removing angular momentum from the gas in the clump (see Section \ref{sec:comp}). Several studies have looked at the collapse of a molecular cloud with a magnetic field aligned with the rotation axis \citep[][]{Allen2003,Machida2005,Galli2006,Price2007,Hennebelle2008,Mellon2008,Duffin2009,Seifried2011,SantosLima2012,Matsumoto2017}. For large values of \muphim($\gtrsim 10$), \ie\ low magnetic field strengths, the impact is minor. However, for stronger magnetic fields the transfer of angular momentum becomes efficient and most workers find that the disk generally fails to form, contrary to observation. This is termed the ``magnetic braking catastrophe."

Before discussing some of the work that has been done to resolve this problem, we note that there are open questions as to the numerical resolution that is required to successfully simulate disk formation. \citet{Joos2012} carried out a number of ideal MHD simulations in which the magnetic field was parallel to the angular momentum of the collapsing gas cloud. They did not use sink particles. For $\muphi=2,\, 3$, they found no disk in their simulation with their standard resolution (0.5 au); however, in a simulation of the $\muphi=2$ case with 1.5 times greater resolution, they found a disk of $0.006 M_\odot$ by the end of the run. On the other hand, for $\muphi=5$, they found a large disk ($M\simeq 0.05 M_\odot$, $R\simeq 150$ au) at their standard resolution, and a somewhat more massive disk when they increased the spatial resolution. For $\muphi=17$, they also found a large disk, suggesting that disks can form for $\muphi\geq 5$. However, when they reduced the time step by a factor 2 for the $\muphi=5$ at the standard resolution, the disk disappeared. The importance of high resolution has also been discussed by \citet{Machida2014} (see below), who emphasized the importance of resolving the non-ideal effects that operate on small scales. \citet{Tomida2015} pointed out the possible importance of the angular momentum lost by the use of sink particles. We note that an upper limit on the momentum lost from the simulation through the use of a sink particle is the stellar mass times the specific angular momentum at the radius of the accretion zone associated with the sink particle, $J_{\rm sink}\sim M_*R_a(GM_*/R_a)^{1/2}$, where $R_a$ is the radius of the accretion zone. By comparison, the angular momentum of the disk is $J_d\sim M_d R_d (GM_*/R_d)^{1/2}$, under the assumption that the disk mass, $M_d$, is small compared to the stellar mass. The ratio of these angular momenta is $J_{\rm sink}/J_d\sim (M_*/M_d)(R_a/R_d)^{1/2}$, which is small only for high resolution, so that $R_a\ll R_d$.

Suggested solutions to the magnetic braking catastrophe generally fall within one of three categories: (1) misalignment between the rotation axis and the magnetic field, (2) non-ideal MHD effects, and (3) turbulence. Late-time accretion has also been suggested: \citet{Mellon2009} pointed out that outflows could remove the envelope that absorbed the angular momentum from the gas accreting onto the protostar. This argues for running the simulations for a long time, as done by \citet{Machida2013}, \citet{Machida2014} and \citet{Tomida2015}, for example, all of whom found substantial disks (see below). However, these simulations are for isolated clouds and do not include possible sinks for the angular momentum of the protostellar envelope.

Misalignment between the rotation axis and the magnetic field was proposed as the mechanism for disk formation by \cite{Hennebelle2009}. Using three dimensional adaptive mesh refinement MHD simulations, these authors showed that even small angles between the rotation axis and the magnetic field can greatly increase the likelihood of forming a disk. In general, the misalignment allows disks to form in environments where \muphim\ is $\sim$3-5, depending on the degree of misalignment \citep{Joos2012}. These authors showed that misalignment significantly reduced the outward flux of angular momentum carried by the field.

\cite{Li2013} showed that the initial conditions adopted for the field affect the outcome. Whereas \citet{Joos2012} assumed that the cloud was initially centrally concentrated and threaded by a uniform field, \citet{Li2013} assumed that the cloud initially had a uniform density. As a result, when the cloud contracted and became centrally concentrated, the field lines spread out and exerted a greater torque than for the case considered by \citet{Joos2012}. The initial condition adopted by \citet{Joos2012} allowed formation of a disk for $\mu_\Phi=3$ at inclinations $\ga 45^\circ$, whereas that adopted by \citet{Li2013} did not.

Non-ideal MHD effects--Ohmic dissipation, ambipolar diffusion (AD), the Hall effect and global reconnection--have for the most part not been shown to permit the formation of large, {\it persistent} disks unless the resistivity \citep{Krasnopolsky2010} or ambipolar diffusivity \citep{Zhao2016} is substantially increased. Axisymmetric simulations including Ohmic dissipation with standard values of the resistivity have found that it is effective in removing excess magnetic flux only at small radii close to the protostar, so it fails to produce any disks larger than about $10\, R_{\odot}$ \citep{Dapp2010,Krasnopolsky2011}.

 Note, however, that if the non-ideal effects are strong enough that disk accretion occurs via numerical viscosity (existing simulations do not include the magnetorotational instability or physical viscosity) rather than magnetic or gravitational torques, then a disk will grow in order to conserve angular momentum.

An extensive set of simulations including Ohmic dissipation for the case of an aligned magnetic field have been carried out by \citet{Machida2014}. They focused on the case $\muphi=3$ and usually used a resolution of 0.4--0.6 au. For the case of a cloud with an initially uniform density, they found that no persistent disk formed in 10 of the 11 cases they considered, consistent with the results of \citet{Li2013}. However, if they set the threshold density for the formation of a sink particle to be very high ($10^{14}$ cm$^{-3}$), then a modest disk ($R\sim 20-30$ au) formed. Since the resistivity scales with the density, this high density enables Ohmic dissipation to be more effective near the protostar. However, it is not clear how a phenomenon that occurs in the inner few au can affect the angular momentum of gas at distances an order of magnitude or more farther out. When they carried out simulations for the aligned case similar to those of \citet{Hennebelle2009} and \citet{Joos2012} (although with a uniform field rather than one corresponding to a spatially constant mass-to-flux ratio), they found similar results: a disk of size $\la 10$ au and mass $\la 0.01 M_\odot$. When Ohmic dissipation was included in this model, the disk was somewhat larger ($\sim 20$ au) and significantly more massive ($\sim 0.2 M_\odot$); however, this case is not very physical, since a star would have formed long before then. Inclusion of a sink particle for this case reduced the disk mass by an order of magnitude. In a third set of simulations, \citet{Machida2014} studied the collapse of Bonnor-Ebert spheres and found large disks, with radii between 50 and 100 au.
It is not clear why the results of this set of simulations differ so much from those based on the initial conditions of \citet{Joos2012}, since the density distribution for the latter ($\rho\propto 1/[1+(r/r_0)^2]$) is quite similar to that of a Bonnor-Ebert sphere \citep{Dapp2009}. One possibility is that the specific angular momentum for the Bonnor-Ebert simulations is about 5 times that for the simulations based on the initial conditions of \citet{Joos2012}. \cite{Tomida2017} used the same code, including Ohmic dissipation, and found the formation of a large, rotationally supported disk with a radius of $\sim$200 au and a mass of 0.2 M$_\odot$. They focused on comparing their results with observation and did not study the role Ohmic dissipation played in the formation of the disk. The simulations of \citet{Machida2013}, \citet{Machida2014} and \citet{Tomida2017} are the only simulations we are aware of in which large disks are produced in non-turbulent simulations in which $\muphi\leq 3$ and in which the angular momentum and the field are aligned.

Several authors have studied the effect of ambipolar diffusion (AD), which is the most important non-ideal MHD process over a significant density range relevant for protostellar disks. Axisymmetric simulations by \citet{Mellon2009}, \citet{Li2011} and \citet{Dapp2012} concluded that AD fails to enable the formation of rotationally supported disks.
Using 2D simulations, \citet{Zhao2016} showed that depletion of very small grains, which is plausible in dense molecular gas, can significantly increase the ambipolar diffusivity and thereby allow the formation of long-lived, rotationally supported disks.

3D simulations by \citet{Duffin2009} found that AD enabled the formation of small disks less than 10 au in size for $\mu_{\Phi}=3.5$. In a recent thorough study, \citet{Masson2016} found somewhat larger disks (15-30 au) for \muphim=2. However, for $\mu_{\Phi}=5$, they found a disk of mass $0.14\, M_{\odot}$ and radius 80 au. It should be noted that 3D simulations with AD are so time consuming that to date they have treated only the formation of the first Larson core \citep{Larson1969}, so such simulations do not provide evidence on the existence of disks around protostars during the main phase of their evolution \citep{Zhao2016}.

\citet{Tsukamoto2015a} found that AD and Ohmic diffusion resulted in the formation of a small ($\sim 1$ au) disk associated with the formation of the first core. \cite{Hennebelle2016} used analytic arguments with AD to suggest that protostellar disks should form with only a weak dependence on the core mass, magnetic field intensity, or turbulence level. They compared their analytic estimate with non-ideal MHD simulations and generally found agreement to within a factor of 2, although in some cases larger disks were found when spiral structure developed due to gravitational instability in the disk. (\citealp{Tomida2017} also found that gravitational torques associated with spiral density waves caused the disk to grow.) We note that the simulations of \citet{Hennebelle2016} all included turbulence, which is discussed below.

There is a controversy associated with the efficacy of the Hall effect enabling disk formation. Using 2D axisymmetric simulations,
\cite{Krasnopolsky2011} found that realistic values for the Hall coefficient were insufficient to lead to the formation of rotationally supported disks. Only an unphysically large Hall coefficient produced a disk with a radius exceeding 10 au, although their initial conditions had $\muphi=1.29$.
\citet{Krasnopolsky2012} confirmed this conclusion with 3D simulations.
\citet{Wurster2016} included all three non-ideal effects in SPMHD simulations and found that the Hall effect was the most effective in fostering the formation of disks. For the parameters they considered, non-ideal MHD approximately doubled the mass and radius of the disks formed under the same conditions with ideal MHD, but it did not enable the formation of a persistent disk when no disk was formed with ideal MHD.
\citet{Tsukamoto2015b}
on the other hand found that a moderate-size disk ($\sim 20$ au) could form when the field and angular momentum were anti-parallel, so that the Hall effect worked to spin up the disk. These simulations were extended to the case of misaligned field and angular momentum by \citet{Tsukamoto2017}
with consistent results.
It should be noted that these simulations are very time consuming and therefore extended only through the formation of the first core; we therefore do not regard these simulations as demonstrating the formation of large, {\it persistent} disks. By contrast, the simulations of \citealp{Krasnopolsky2011,Krasnopolsky2012}
and of \citet{Li2011} did not attempt to follow processes inside a radius of 6.7 au and so were able to carry out their simulations to much later times. Their conclusion that the Hall effect does not allow disk formation applies to the case in which a significant fraction of the mass of the cloud had accreted onto the central star. However, they were unable to address the question of whether the disk formed with the first core could evolve into a persistent disk, as suggested by \citealp{Tsukamoto2015b,Tsukamoto2017} 
The question of whether the Hall effect permits the formation of a large, persistent circumstellar disk thus remains unresolved.

Finally, we note that global reconnection (i.e., a change in the topology of the field that results in the field in the inner disk becoming disconnected from the envelope) has been mentioned as possibly influencing the formation of disks \citep{Li2014b,Gonzalez2016}, but it is not strong enough to enable the formation of a disk in the absence of misalignment and turbulence \citep{Li2014b}.

The conclusion from most of these studies is that in the absence of turbulence, large ($\sim 100$ au), persistent protostellar disks cannot form in regions in which the mass-to-flux ratio has its typical value, $\mu_\Phi\simeq 3$, even allowing for non-ideal MHD, unless the magnetic field is sufficiently misaligned with the angular momentum. In fact, \citet{Li2014b} have argued that in ideal MHD, the minimum value of \muphim\ required for disk formation is $\ga 4$, even allowing for disk misalignment. On the other hand, simulations that include turbulence always result in reasonably large disks, even for $\muphi=2$ and ideal MHD, provided that the resolution is sufficient ($\Delta x\la 1$ au) \citep{SantosLima2012,Seifried2012,Joos2013,Myers2013}. It is natural that star-formation simulations should include turbulence since it appears to be ubiquitous in star-forming regions: When mapped with molecular tracers such as CO, CS, and C$^{18}$O, it appears that the internal motions in these cores are both supersonic and turbulent \citep[\eg][]{Barranco1998,Goodman1998,Caselli2002,Tafalla2004,Miettinen2010,Sanchez2013}. Turbulence can promote the formation of protostellar disks in several ways \citep{Seifried2012}:

\begin{itemize}
\item[1)] Turbulent misalignment. Turbulence naturally creates a time dependent misalignment between the angular momentum and magnetic field that promotes disk formation.
\item[2)] Turbulence-weakened magnetic torque: Turbulence prevents the formation of the ordered toroidal field that is responsible for extracting angular momentum from the accreting gas.
\item[3)] Turbulent reconnection/turbulent resistivity/turbulent magnetic diffusivity that weakens the field.
\end{itemize}

There is general agreement that turbulent misalignment is not the dominant effect in allowing disk formation: By itself, turbulent misalignment would presumably allow disk formation only under conditions in which misaligned disks form in the absence of turbulence, whereas in fact turbulence increases the range of conditions under which disks form. Calculations have shown that both misalignment \citep{Joos2012} and turbulence \citep{Seifried2012} foster disk formation by reducing the magnetic torque on the accreting gas, and \citet{Joos2013} showed that turbulence reduced the magnetic torque more than did misalignment.

\citet{SantosLima2012} have proposed that turbulent reconnection (also termed ``reconnection diffusion") can account for disk formation. This process naturally occurs in turbulent, magnetized media, as described in the theory of \citet{Lazarian1999}. While in principle turbulent reconnection can be distinguished from a generic turbulent resistivity or turbulent magnetic diffusivity by its particular properties--for example, its rate is proportional to $\ma^3$ for $\ma\la 1$, where $\ma$ is the \alfven\ Mach number--no tests to date have identified turbulent reconnection as the specific process that aids disk formation. In their simulation of accretion of magnetized gas onto a protostar of mass $0.5 M_\odot$, \citet{SantosLima2012} injected turbulence with a turbulent diffusivity $\eta_{\rm turb}\sim L_{\rm inj}v_{\rm turb}\sim 10^{20}$~cm$^2$~s. They showed that this turbulence reduced the magnitude of the magnetic field within 400 au of the star by an amount comparable to that in a non-turbulent simulation with an electrical resistivity of about the same magnitude, which is more than 100 times the expected value and large enough to permit the formation of large disks according to \citet{Krasnopolsky2010}.

In their simulations of disk formation, \citet{Joos2013} found that the mass-to-flux ratio \muphim\ increased above its initial value in the vicinity of the star and attributed this to turbulent magnetic diffusivity. They regard turbulent magnetic diffusivity as an important contributor to disk formation, but not sufficient by itself.
{\color{black}
\citet{Kuffmeier2016} studied the formation of protoplanetary disks using a high-resolution zoom-in simulation. Their simulation is unique in that it had an outer scale of 40 pc, and used adaptive mesh refinement to achieve a highest resolution of 2 au around 9 protostars. Turbulence was first driven artificially and then by simulated supernovae, such that the velocity dispersion agreed approximately with that found in the solar neighborhood by \citet{Larson1981}. The cloud was initially weakly magnetized, with \muphim\ $\sim$ 7.5; on scales of 10$^3$-10$^4$ au, \muphim\ was in the range 2-10. They found that substantial disks formed, often with large misalignments, around a majority of the 9 protostars they studied. Their results show that \muphim\ increased in the vicinity of the stars, being generally above 5 inside 1000 au and significantly above the initial value of 7.5 inside 100 au.}

As noted by \citet{Seifried2012,Seifried2013}, however, weakening the magnetic field within $\sim 100$ au of the protostar is ineffective at forming a disk at that radius. To make this argument more quantitative, consider a centrally condensed ($\rho\propto r^{-2}$) protostellar core of radius 0.1 pc that is rotating uniformly with $\bobs=0.02$, the typical observed value. The half-mass point is at $r_i\simeq 10^4$ au. Equation (\ref{eq:rcf}) implies that in the absence of any torques, gravitational collapse would lead to a disk of radius about 600 au at that time that half the mass has collapsed into a protostar and disk. If torques acting beyond 600 au had removed 75\% of the angular momentum, then the outer radius of the disk would be only about 40 au. In order to enable the formation of a Keplerian disk with a radius $\sim 100$ au, magnetic torques would have to be suppressed at radii beyond 600 au. It appears then that while turbulent reconnection or turbulent magnetic diffusivity can aid disk formation (and persistence), it alone cannot account for the reduction in magnetic braking necessary for the formation of large Keplerian disks.

In this paper we carry out numerical experiments aimed at increasing our understanding of the processes involved in creating protostellar disks. In particular, in order to determine the effect of turbulent misalignment, we have developed a procedure that ensures that the misalignment angle between turbulence and the magnetic field is initially very small throughout the whole cloud. Our simulations use ideal MHD; as discussed above, non-ideal MHD alone is not able to produce disks of the size observed. However, our sink particles include one of the major effects of non-ideal MHD since they do not accumulate any magnetic flux (Section \ref{sec:num}), thereby ensuring agreement with the observation that the magnetic flux trapped in stars is a negligible fraction of that initially associated with the gas in the star (e.g., \citealp{McKee2007}). In this respect, our simulations differ from those by \citet{Li2014b}, which are truly ideal MHD and accumulate flux inside the accretion radius.

Several different initial conditions have been used in the study of disk formation: (1) a uniform field in a centrally concentrated cloud (e.g., \citealp{Machida2014}), (2) a cloud and magnetic field that are both centrally concentrated (e.g., \citealp{Myers2013}), (3) a constant or approximately constant mass-to-flux ratio in a centrally concentrated cloud (e.g., \citealp{Hennebelle2009,Seifried2012}), and (4) a uniform field in a uniform cloud (e.g., \citealp{Li2013}). It is not obvious how the first condition would arise in nature. The interpretation of the results for (1) and (2) is complicated by the fact that \muphim\ increases toward the center of the cloud \citep{Li2013,Myers2013}, which fosters disk formation at early times. For example, the profiles used in \cite{Myers2013} give $\muphi=5.6$ for the central flux tube, compared to $\muphi=2$ for the cloud as a whole. The third and fourth initial conditions both have approximately constant values of $\muphi(r)$ (for the uniform cloud, the central value of \muphim\ is 1.5 times the value for the whole cloud). Both initial conditions assume that the field is initially straight. It is not clear which of these is a closer approximation to reality in a turbulent medium. In this paper we adopt the third initial condition and enforce a constant value of $\muphi(r)$. A consequence of our assumption is that the field becomes very small on the circumference of the cloud in the plane normal to the field; in our implementation it is larger than the field in the diffuse interstellar medium.

Finally, our simulations focus on large clouds, with masses of $300M_\odot$. This is somewhat more realistic than considering the collapse of isolated clouds of much smaller mass, but star-disk interactions do introduce a complication in the analysis. This paper is organized as follows. In \S 2 we describe the numerical setup and initial conditions. In \S 3 we present the results of our simulations focusing on the conditions in which protostellar disks form and evolve. We discuss the impact and evolution of \muphim\ in \S4. In \S5 we compare our results to previous published works. Finally, concluding remarks and discussion are given in \S 6.

\section{Numerical Setup}
\label{sec:num}
All the simulations presented here were performed using {\rm $ ORION2$ } \citep{LiPS2012}, our adaptive mesh refinement magneto-radiation-hydrodynamics code. The particulars of the code have been discussed in previous articles \citep[\eg][]{Klein1999,Krumholz2007,Myers2011,Cunningham2011,Myers2011,LiPS2012,Myers2013,Lee2014} and we refer interested reader there for the specific numerical methods and algorithms used within ORION2.

\begin{table*}
\resizebox{\linewidth}{!}{
\centering
\begin{threeparttable}
\caption{Simulation Runs}
\begin{tabular}{lccccccccccc}
\hline
Name  & Mass & \muphim & $\theta$ & $\Delta x_{\ell_{\max}}$ & ${\frac{E_{\rm Rot}}{E_{\rm KE}}}$ & \alphavirm & $\beta_{\rm obs}$ & $\calm$ & $\ma$ & N$_{\rm star}$ & N$_{\rm disk}$ \\
\hline
\hline
AT0       & 300 & 2.0   & 0.0      & 2.50     & 0.595       & 0.071 & 0.0212 & 2.8  & 0.3 & 0  & 0 \\
AT10      & 300 & 2.0   & 0.0      & 2.50     & 0.232       & 0.302 & 0.0350 & 5.7  & 0.6 & 1  & 0 \\
AT50      & 300 & 2.0   & 0.0      & 2.50     & 0.103       & 1.101 & 0.0567 & 11.0 & 1.1 & 4  & 0 \\
AT100     & 300 & 2.0   & 0.0      & 2.50     & 0.074       & 2.074 & 0.0765 & 15.0 & 1.6 & 16 & 0 \\
AT100L8   & 300 & 2.0   & 0.0      & 1.25     & 0.074       & 2.074 & 0.0765 & 15.0 & 1.6 & 6  & 0 \\
AT100NR   & 300 & 2.0   & 0.0      & 2.50     & 0.051       & 2.072 & 0.0524 & 15.0 & 1.6 & 12 & 0 \\
RT100TH37  & 300 & 2.0   & 37.0    & 2.50     & 0.051       & 2.090 & 0.0524 & 15.0 & 1.6 & 10 & 1 \\
RT100TH0   & 300 & 2.0   & 0.0     & 2.50     & 0.013       & 2.071 & 0.0139 & 15.0 & 1.6 &  3 & 1 \\
\hline
RT10M5    & 5   & 2.0   & 44.0     & 1.75     & 0.012       & 0.177 & 0.0006 & 0.9  & 0.5 & - & - \\
RT10M5MU5 & 5   & 5.0   & 44.0     & 1.75     & 0.012       & 0.177 & 0.0006 & 0.9  & 1.2 & - & - \\
RT10M5L8  & 5   & 2.0   & 44.0     & 0.87     & 0.012       & 0.177 & 0.0006 & 0.9  & 0.5 & - & - \\
AT10M5	  & 5   & 2.0   & 0.0      & 1.75     & 0.004       & 0.187 & 0.0002 & 0.9  & 0.5 & - & - \\
\hline
\end{tabular}
\label{tab:simlist}
\begin{tablenotes}{}{
\item \textbf{Note:}
Names: Runs labeled AT have the turbulence aligned with the initial magnetic field; RT denotes random turbulence. The number after AT or NT denotes the percentage ratio of the virial parameter to about 2.1. All but two of the simulations have 7 levels of refinement; the two with eight levels are labeled L8. Mass is the initial mass of the cloud in units of \msun\ (runs in which the initial mass is $5\ M_\odot$ are labeled $"M5"$), $\Delta x_{\ell_{\max}}$ is the finest resolution in units of au, \muphim\ is the mass-to-flux ratio (one run has $\muphi=5$, and that is labeled \'MU5\'), $\theta$ is the initial misalignment angle, ${{E_{\rm Rot}}/{E_{\rm KE}}}$ is the ratio of the rotational energy to kinetic energy, $\alpha_{\rm vir}$ is the initial virial parameter, and $\beta_{\rm obs}$ is the observational ratio of rotational and gravitational energy. $\calm$ and $\ma$ are the initial Mach and Alf{\'v}en Mach numbers, respectively. N$_{\rm star}$ gives the number of star particles formed, while N$_{\rm disk}$ gives the number of rotationally supported disks formed. The velocity field in all models includes both a turbulent component and a solid body component, except for AT100NR, which includes only the turbulent component, and AT0, which contains only a solid body component aligned with the field. RT100TH37 is a completely random model in which the net turbulent angular momentum is at an angle of 37\deg\ with respect to the initial magnetic field, whereas RT100TH0 is initially aligned using a single shell, see Appendix~\ref{turbalign}. Due to the stiffer equation of state used in the $"M5"$ runs, no star particles are formed.
}
\end{tablenotes}
\end{threeparttable}
}
\end{table*}

\subsection{Initial Conditions}
\label{sec:ics}
We closely follow the initial setup presented in \cite{Myers2011,Myers2013} and \cite{Cunningham2011} except for the initial magnetic field profile and initial turbulent velocity field. In most runs, the cloud is initialized as an isolated sphere with a total mass of 300 \msun, radius $R_c$=0.1 pc, and temperature $T_c$=20 K. We also run several smaller mass clouds to compare with \cite{Joos2013}; these are described in \S \ref{sec:prevstudies}. The density follows a power-law profile proportional to $r^{-1.5}$, which is consistent with models \citep{McKee2002,McKee2003} and observations \citep{Beuther2007}. The density at the edge of the core is given by
\be
\rho_{\rm edge} = \frac{3 M_c}{8 \pi R_c^3}.
\ee
To prevent unphysically large densities at the center of the cloud, we use a parabolic density profile ($\rho(r) \propto (1-(r/r_0)^2)$) for the interior 5\% of the cloud. The mean density of our cores is $\bar{\rho}\approx$4.8$\times$10$^{-18}$ g cm$^{-3}$, or $\bar{n}_{\rm H}$=2.4$\times$10$^{6}$ H nuclei cm$^{-3}$, assuming an average molecular weight of 2.33 $m_{\rm H}$. This value for the mean density sets the characteristic timescale for gravitational collapse, given by
\be
t_{\rm ff} = \sqrt{ \frac{3 \pi}{32 G \bar{\rho}}} \approx {\rm 30.2 \ kyr}.
\ee
The surface density of the cores, $\Sigma_c = M_c / \pi R_c^2 \approx 2.0$ g cm$^{-2}$, is adopted to approximate those observed in galactic regions of high-mass star formation. \cite{McKee2003}, for example, inferred a mean $\Sigma \sim$ 1 g cm$^{-2}$ from a sample of high-mass clumps in \cite{Plume1997}.

All of the simulations presented here assume a barotropic equation of state. We have constructed a piecewise equation of state based on the results of \cite{Banerjee2006}. There the authors give the effective adiabatic index, $\gamma$, as a function of density; dust cooling and H$_2$ dissociation are included. Using a constant average mass per particle of $2.33 m_{\rm H}$ gives the following functional form for the pressure:
\begin{align}
&P=\rho c_{s,0}^2      & & \rho < \rho_c  \\
&P=\rho_c c_{s,0}^2 \left({\rho}/{\rho_c}\right)^{1.1}       & & \rho_c< \rho < \rho_d  \nonumber \\
&P=\rho_c c_{s,0}^2 \left({\rho_d}/{\rho_c}\right)^{1.1} \left({\rho}/{\rho_d}\right)^{4/3}       & & \rho_d < \rho < \rho_e  \nonumber \\
&P=\rho_c c_{s,0}^2 \left({\rho_d}/{\rho_c}\right)^{1.1} \left({\rho_e}/{\rho_d}\right)^{4/3} \left({\rho}/{\rho_e}\right)^{1.0}        & & \rho_e < \rho, \nonumber
\end{align}
where $\rho_c,\rho_d,\rho_e$ are 5$\times$10$^{-15}$, 2$\times$10$^{-13}$, 2$\times$10$^{-8}$ g cm$^{-3}$, respectively, $\rho$ is the mass density, and $c_{s,0}$ is the initial sound speed of 0.266 km s$^{-1}$, corresponding to an initial temperature of $T_c$=20 K.

Each core is placed in at the center of a cubic box with side length equal to $4R_c$, so that the sides of the domain are far enough from the cloud that interactions between the cloud and the boundaries are minimized. We use outflow boundary conditions for the MHD variables, which requires the gradients for $\rho$, $\rho {\bf v}$, $E$, and ${\bf B}$ to equal zero at the edge of the domain. In addition, we require that the gravitational potential be zero at the boundaries.

Following the initial conditions of \cite{Seifried2012}, we initialize the velocity field as composed from both turbulence and solid body rotation. The solid-body rotation is aligned with the initial magnetic field, with a $\beta_O$ = 0.02. The turbulent velocities are drawn from a Gaussian random field with a power spectrum of $P(k) \propto k^{-2}$, appropriate for the highly supersonic turbulence found in molecular cloud clumps. The perturbations are added using the following steps: first, we generate a 256$^3$ perturbation cube using the method of \cite{Dubinski1995} with power on scales ranging from $k_{\min}$=1 to $k_{\max}$=128. This perturbation cube is then placed over the simulation volume and, where needed, the perturbations are either coarsened or interpolated so that we can represent the perturbations at all levels of refinement. The perturbations are then scaled by a perturbation velocity to put the cloud at the target $\alpha_{\rm vir}$.  For example, following \cite{Myers2013}, we initialize our fiducial cloud with a one-dimensional velocity dispersion of $\sigma_c$=2.3 km s$^{-1}$, chosen to put the core into approximate virial balance. The virial parameter $\alpha_{\rm vir}\equiv5\sigma_c^2R_c/GM_c$ \citep{Bertoldi1992} is defined so that a spherical cloud of constant density has $\alphavir=1$ when it is in virial equilibrium. Our choice of parameters gives $\alphavir \approx 2.1$ , which gives slightly more kinetic energy than gravitational energy in our cores. This is justified since no additional driving is performed once the simulation is started, which allows the turbulence to decay. For example our fiducial model with an initial \alphavirm=2.1 has decayed to \alphavirm=1.7 by the end of the simulation.

These perturbations impart angular momentum into the cloud, which generates a natural misalignment between the initial magnetic field, ${\bf B_0}$, which is always aligned with the $z$-axis, and the total angular momentum, ${\bf J}$. Previous studies have shown that even a slight misalignment ($\sim$10$^{\circ}$) between ${\bf B}$ and ${\bf J}$ has a profound impact of the formation of disks \citep[\eg][]{Hennebelle2009,Joos2012,Seifried2012,Li2013}. In fact, in all the simulations presented by these authors ${\bf B_0}$ and ${\bf J}$ are initially misaligned. The models presented in \cite{Joos2012}, for example, have initial misalignments between 40-60$^{\circ}$.

To determine the importance of the turbulence without the accompanying misalignment, we process the perturbation cube such that the angular momentum from turbulence is aligned with the magnetic field; for details see Appendix \ref{turbalign}. With our procedure, we find that the misalignment between the total angular momentum vector and the magnetic field is much less than one degree for the whole cloud and $\approx$1.86$^{\circ}$ for the central core.  We make no effort to filter out compressive modes in the initial velocity field. The final perturbation cube has a solenoidal forcing to total forcing ratio of 0.58, close to the expected theoretical value of 2/3 \citep{Federrath2010}.

We have also included the ability to rotate the perturbations such that they align with any arbitrary angle in the $x-z$ plane, aligning them with a given axis $\hat{a}$. Any initial misalignment, given by $\theta$, can be achieved by a simple rotation about the $y$-axis.

When $\theta = 0$, $\hat{a}\equiv\hat{z}$ and $\theta=90, \hat{a}\equiv\hat{x}$. The solid body rotation velocities are initialized as $v_x=-\Omega {y}$ and $v_y=\Omega {x}$, where $\Omega$ is a constant such that the ratio of the rotational energy and gravitationally potential energy is $\beta_O$=0.02, consistent with observations of molecular cores \citep{Goodman1998}. These can then be rotated by $\theta$ to align with $\hat{a}$.

The importance of the magnetic field is quantified by the normalized mass-to-flux ratio, $\muphi=M/M_\Phi$, where
\be
M_\Phi=c_\Phi\frac{\Phi}{\sqrt G}
\label{eqn:cmass}
\ee
is the magnetic critical mass, $\Phi$ is the magnetic flux threading the cloud, and $c_\Phi$ is a numerical constant. Cores are magnetically supported against collapse when $\muphi<1$ whereas cores with $\muphi>1$ can collapse provided gravity is strong enough to overcome other sources of support. For the case of a cloud that is initially spherical and threaded by a uniform field, $c_\Phi= 0.126$ \citep{Mouschovias1976}. For the case in which the field threading the initial cloud has a constant mass-to-flux ratio, $c_\Phi= 0.17$ \citep{Tomisaka1988}, which to within the accuracy of the calculations agrees with the value $c_\Phi=1/(2\pi)$ for a uniformly magnetized sheet \citep{Nakano1978}.

Since our simulation has an initially uniform mass-to-flux ratio, we adopt this value so that
\be
\mu_{\Phi} = 2\pi\sqrt G \left(\frac{M}{\Phi}\right).
\label{eqn:muphi}
\ee
Denoting the value of the normalized mass-to-flux ratio calculated using $c_\Phi=0.126$ by $\mu_{\Phi'}$, we have $\muphi=0.79\mu_{\Phi}'$.

For our simulations, we adopt $\muphi=2$, the value generally quoted based on Zeeman splitting measurements of CN \citep{Falgarone2008}, which probes densities of $10^{4}\la \nh<10^{6}$ cm$^{-3}$, and of OH \citep{Troland2008}, which probes densities of $10^{3}<\nh<10^{4}$ cm$^{-3}$.
It should be noted that \cite{Li2015} have shown that the median value of \muphim\ in the sample of molecular clumps analyzed by \cite{Crutcher2010} is actually about 3, somewhat larger than previously thought because the observational value is density weighted whereas the value from theory is not. Observational difficulties measuring the magnetic field strength can lead to a substantial scatter in \muphim\ \citep[\eg][]{Crutcher2012, Li2012}. The cores we simulate thus have mass-to-flux ratios somewhat lower than the typical value, but well within the range of observed values. We require that the magnetic field have a profile that gives a spatially uniform \muphim, similar to the profiles presented in \citet[][]{Joos2012,Joos2013}.

Details of the procedure used to generate the field, which is in the $z$-direction, are given in Appendix~\ref{calcmuphi}.  On the central flux tube, the field is 14 \mig. Near the circumference in the $x-y$ plane the field rapidly drops by more than an order of magnitude to 1.6 \mig; outside the cloud, we allow the field to drop gradually to about 6 \mug, which corresponds to the value in the diffuse ISM \citep{Heiles2005}.

The simulation domain is a cube with side of length $L_0$ and split into $N_0$ cells such that the coarse grid resolution in the base grid is $\Delta x_0=L_0/N_0$.  ORION2 operates within an AMR framework that automatically adds or removes levels of refinement as the simulation develops, up to a maximum level $\ell_{\rm max}$. With $\ell_{\rm max}$ levels of refinement and a refinement ratio of 2, the resolution at the finest level is $x_{\ell_{\rm max}}=\Delta x_0/2^{\ell_{\rm max}}$. The finest resolution for each model is given in Table.~\ref{tab:simlist}. Refinement is based solely on the local magnetic Jeans density, given by \citet{Myers2013}
\be
\rho_{\max} = \frac{ \pi J_{\max}^2 c^2_s } { G \Delta x_l^2} \left( 1 + \frac{ 0.74 } {\beta} \right),
\ee
where $J_{\rm max}$(=0.125) is the maximum allowed number of Jeans lengths per cell, $c_s$ is the local isothermal sound speed, $\Delta x_l$ is the cell size at level $l$, and $\beta=8\pi \rho c^2_s / B^2$ (see \citealp{Myers2013} for details). We also ensure that cells within 16$\Delta x_{\ell_{\rm max}}$ of the star particle are fully refined. This is done recursively up to the highest refinement level until the maximum level is reached. At this point, whenever a cell exceeds the Jeans mass, a star particle is formed, which incorporates the excess mass. If a particle is formed within the accretion region of an adjacent particle, we merge these two particles into a single composite particle. However, we allow the particles to merge only if one of particles is below a cutoff mass, $M_{\rm cutoff}$. If two particles above the cutoff mass get near each other, they are prevented from merging and instead evolve as two separate particles. This prevents unphysical merging of large particles if they get too close to one another \citep{Krumholz2004}. In the simulations presented here, the cutoff is set to $M_{\rm cutoff}$ = 0.05 \msun. This cutoff is justified as this represents the mass at which the `second collapse' to protostellar densities occurs \cite{Masunaga2000}.

Our sink particles are endowed with a number of stellar properties, so we refer to them as star particles. The mass, position, linear momentum, and angular momentum (see \citealp{Fielding2015} for details) of each star particle is evolved along with the fluid. We model mass accretion onto the star particle by including a mass sink term inside a radius, $r_{\rm acc}$= 4$\Delta x$, equal to four grid cells on the finest AMR grid \citep{Cunningham2012,Lee2014}. Although the star-particle creation algorithm was recently extended to include the effects of MHD \citep[\eg][]{Myers2013}, the star particle itself does not have any magnetic flux.  This behavior is consistent with observations showing that the magnetic flux in young stellar objects is orders of magnitude lower than that in the gas that formed these objects. Flux accretion is thus very inefficient, presumably due to non-ideal MHD effects including AD and reconnection \citep{McKee2007,Braithewaite2012,Masson2016}.  As a result, the magnetic field becomes detached from the gas that is accreted onto a star particles and is buoyantly transported away via interchange instabilities. \citet{Masson2016} have pointed out that these interchange instabilities are spurious, and can be avoided by including AD and not using sink particles. Unfortunately, this procedure works only through the phase of the first Larson core \citep{Machida2014}. In our case, since the interchange instability is initiated at the accretion radius, which is resolved by only four cells, numerical diffusion will smooth the field and gas so as to mimic the effects of AD. Our sink particle algorithm should therefore be satisfactory for determining whether large disks can form.

\section{Results}
\label{sec:results}

Table~\ref{tab:simlist} summarizes the models presented here. All our models are first labeled whether the turbulence is initially aligned with the magnetic field or not (AT for aligned turbulence and RT for random turbulence), the initial amount of turbulence (given as a percentage, where 100 corresponds to an initial $\alphavir\sim2.1$), whether the solid body rotation component is included (NR for no solid body rotation), the initial mass of the cloud (M5 for $5\,M_\odot$ clouds), the number of levels of refinement (L8 for 8 levels), and the initial mass-to-flux ratio of the cloud (MU5 for \muphim=5). RT100TH0 is produced by using a single shell in our alignment procedure. That is, at R$_c$ the angular momentum is aligned with B$_0$, but not guaranteed at any other radius, contrary to our AT models.

For brevity we have dropped parts of the name if they are the default values-- $M=300\,M_\odot$, 7 levels of refinement, and \muphim=2. For example, AT100 is our fiducial model and AT100L8 is the same as our fiducial model with one more level of refinement. We ran the 5 \msun\ models in order to compare with \cite{Joos2013}. These models are labeled ``M5" and are described in \S\ref{sec:prevstudies}.

\subsection{Criteria for a Keplerian Disk}
\label{sec:crit}

In an effort to quantitatively study and compare the disks formed in each of our simulations, we follow a procedure similar to the one outlined in \cite{Joos2012} to define our disks. A sphere of radius $r$ is used to calculate the angular momentum vector around the protostar. We then define a cylindrical coordinate system with radius $r$ and height $r/2$ centered on the protostar with the $z$-axis of the cylinder parallel to the angular momentum vector.

We define the disk as a proper Keplerian disk if it meets the following criteria:
\begin{enumerate}
\item We expect that the radial velocity is much smaller than the tangential velocity ($v_{\phi}>f_{\rm threshold}v_r$).
\item The disk is in approximate hydrostatic equilibrium so that the tangential velocity is greater than the vertical velocity ($v_{\phi}>f_{\rm threshold}v_z$).
\item We ensure that the disk is rotationally supported and not thermally supported ($\rho v_{\phi}^2/2>f_{\rm threshold}P_{\rm thermal}$).
\item We visually inspect each candidate disk to ensure that each segment of the disk lies on a common equatorial plane.
\item We use a density cutoff criterion ($\nh>10^9$ cm$^{-3}$) to avoid any large spiral arms and obtain a realistic estimate of the shape of the disk.
\item Finally, we ensure that the velocity profile, \ie v$_{\phi}$ versus radius, of the disk follows a Keplerian profile normalized using the protostellar mass.
\end{enumerate}

Again, following \cite{Joos2012}, we use a $f_{\rm threshold}=2$ for the above criteria. For criterion 4, we ensure that the disk lies within a few cells of the equatorial plane, although in practice little deviation is found. We also find that for criterion 5 $\nh>10^9$ cm$^{-3}$ is sufficient to contain most, if not all, the gas in the disk. Finally, criterion 6 is only valid as long as the stellar mass is greater than the disk mass. We find for the sample of stars shown below that the median (average) of the ratio of the gas mass to stellar mass is $\sim$ 0.05 (0.163) at 50 au and 0.25 (0.43) at 100 au, respectively.

Using this set of criteria, we check for a disk at several radii around each star particle. The physical size of the disk is then determined as the maximum radius where all criteria are met. The radius range is chosen between $r_{\rm min}$=8$\Delta x_{\ell_{\rm max}}$, slightly larger than the accretion radius of the particle, and $r_{\rm max}$=100 au. Each simulation is evolved for $0.3\tff$, after which each star greater than $0.05M_{\odot}$ is analyzed for the existence of a disk given the above criteria. The mass cutoff is chosen because stellar mergers are forbidden once a star reached $M_{*}>0.05\,M_{\sun}$.

\subsection{Large Scale Evolution}

\begin{figure*}
\centering
\includegraphics[trim=0.0mm 0.0mm 0.0mm 0.0mm, clip, scale=0.70]{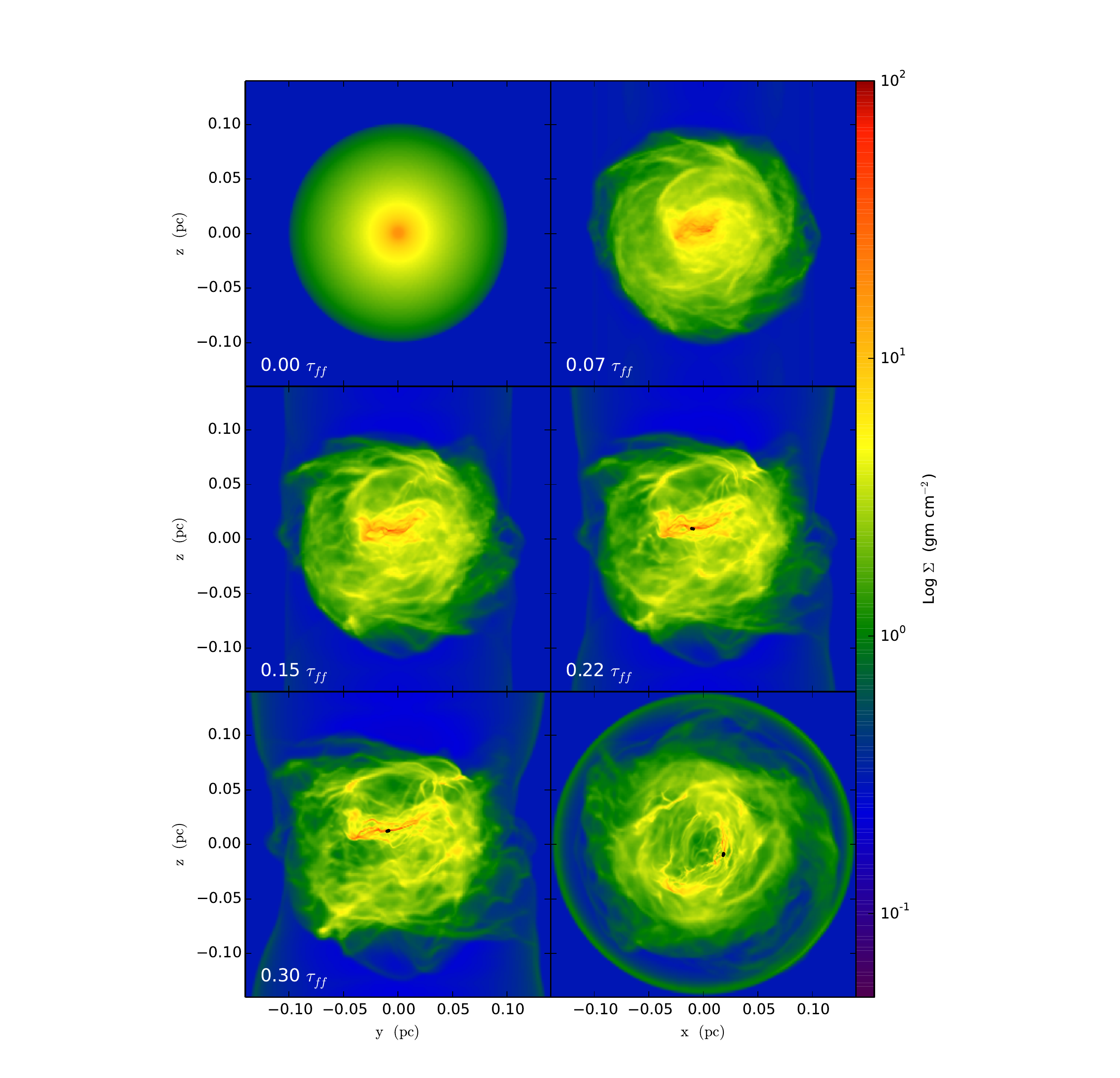}
\caption{Column density projection through the molecular cloud at five times during the evolution of AT100. The projections are taken along the $x$-axis. The time of each projection is given in the label of each panel. The bottom right panel shows a projection along $z$-axis for the final time. Particles are shown as the black points. Note that only 7 stars have formed by $\tau_{ff}$=0.3, all within a small clump. }
\label{fig:gevo}
\end{figure*}

We first consider the large-scale structure of our fiducial model AT100, shown in Figure~\ref{fig:gevo}. As the simulation evolves, the imposed velocity field quickly begins to form several low density filaments. By 0.15\tffm\ a single dominant filament forms near the center of the cloud. Although the initial thermal Mach number of the cloud is large, with $\calm\sim15$, the cloud is only marginally super-\alfvenic, with an \alfven\ Mach number of $\ma\sim$1.6. The existence of such large magnetic signal speeds means that what would be strong shocks parallel to the magnetic field lines would be only weak shocks, or not shocks at all, in flows perpendicular to the magnetic field. The large magnetic field also forces the filament to accrete primarily from along the magnetic field lines. As can be seen in Figure~\ref{fig:gevo}, the filament is well defined along the $y$-axis in the $x$-projections, while very little filamentary structure is seen along the initial magnetic field axis. Thus, the filament is approximately perpendicular to the magnetic field.

\begin{figure*}
\centering
\includegraphics[trim=0.0mm 60.0mm 0.0mm 60.0mm, clip, scale=0.70]{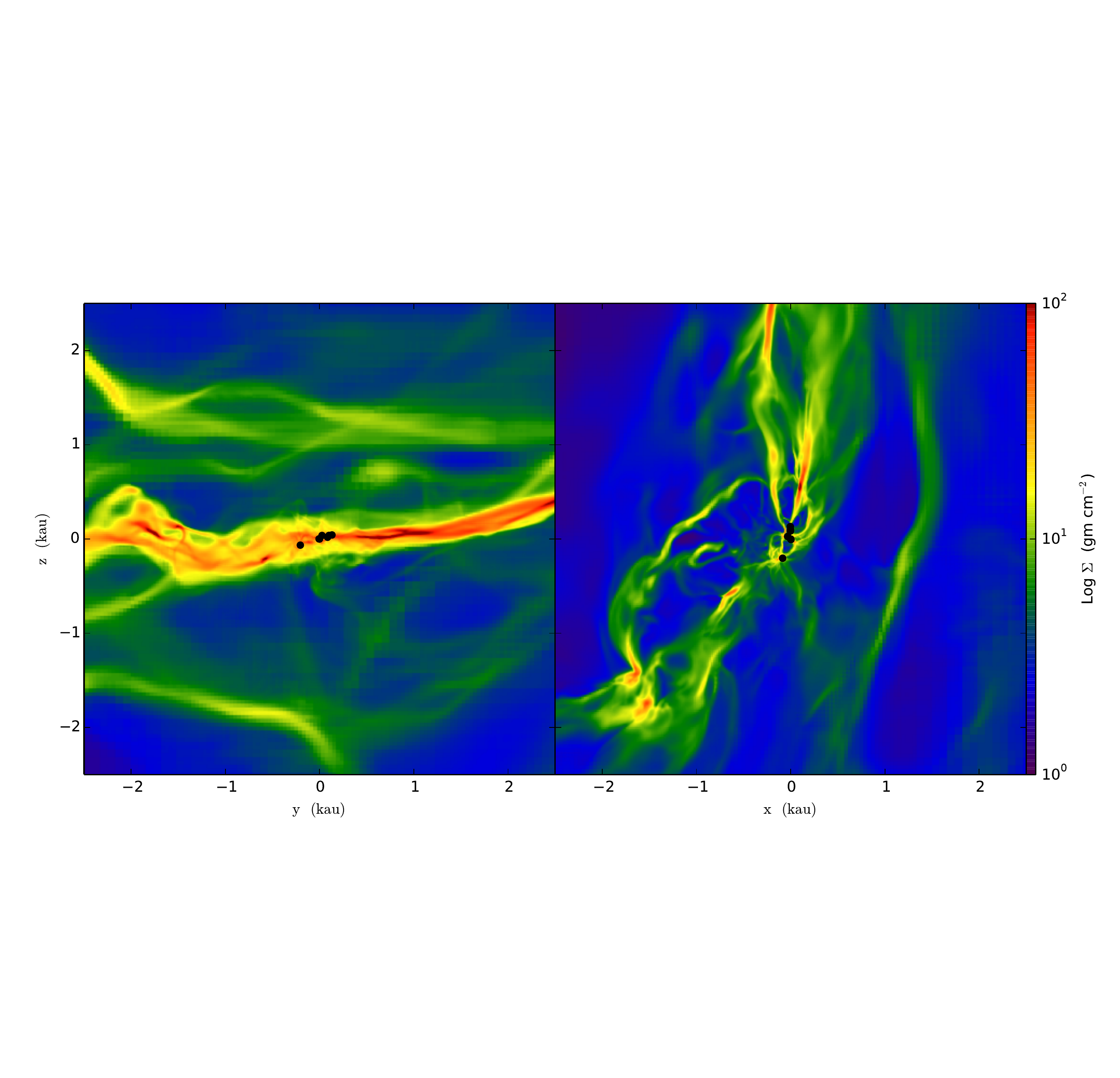}
\caption{Column density projection centered on a region 5000 au in size around the largest particle formed in AT100 at 0.3 $\tau_{ff}$. Particles are shown as the black points.}
\label{fig:gevoz}
\end{figure*}

\subsection{Stellar Properties}

\begin{figure}
\centering
\includegraphics[trim=15.0mm 0.0mm 0.0mm 0.0mm, clip, scale=0.30]{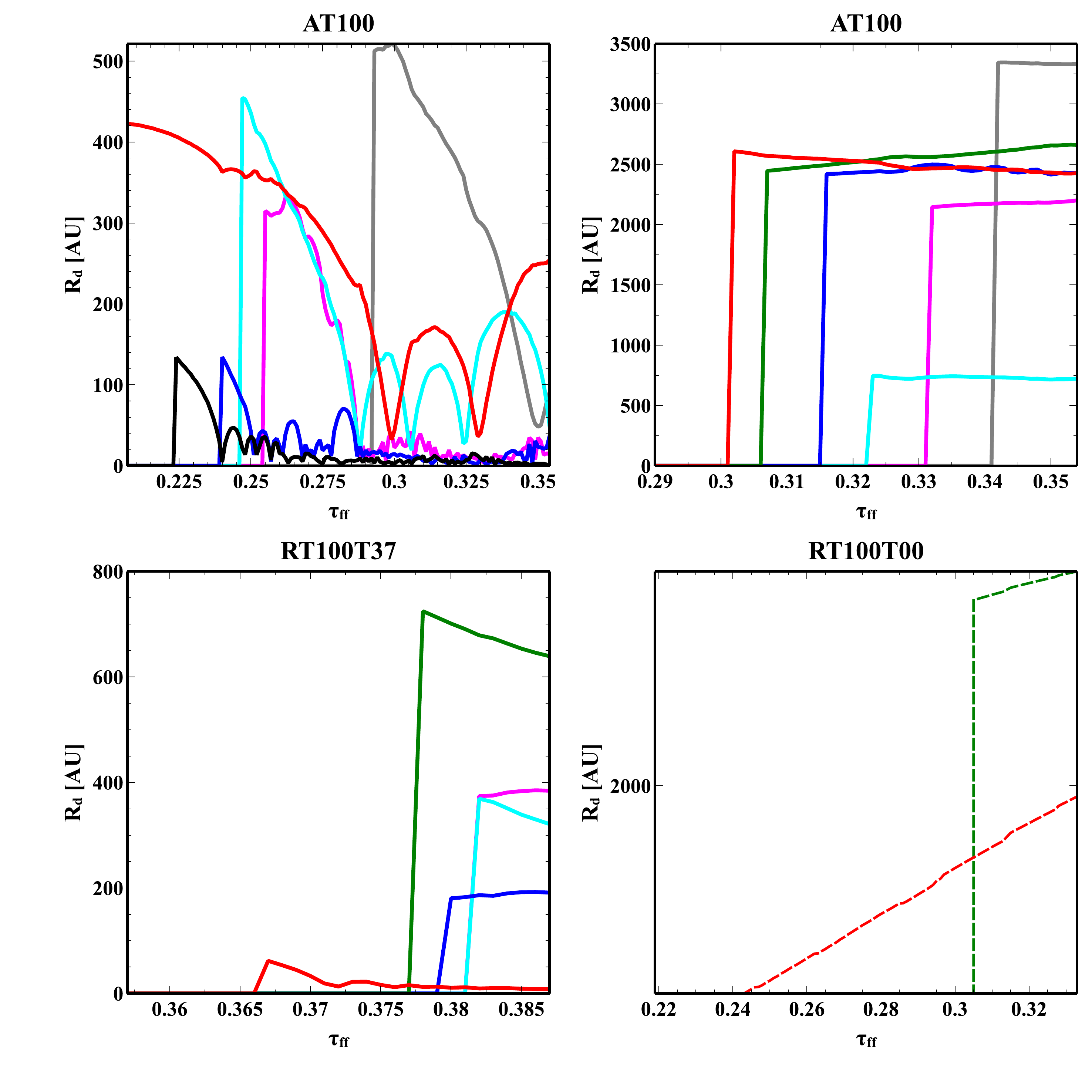}
\caption{Plots of separation distances for {\it Top Left} stars near the central star in AT100, {Top Right:} the remaining 6 stars in AT100, {\it Bottom Left:} RT100TH37, and {\it Bottom Right:} RT100TH0. Only stars greater than the cutoff mass are considered.}
\label{fig:stars}
\end{figure}

We now consider the properties of the stars formed in our simulations.

As the primary filament collapses, star particles begin to form, shown as the black points in Figure~\ref{fig:gevo}. These stars preferentially form in clumps and not spread throughout the filament, as shown in Figure~\ref{fig:gevoz} which zooms in on a region 5000 au in size around the largest particle formed in AT100 at 0.3$\tau_{ff}$. Note that 7 stars have formed in AT100 by $\tau_{ff}$, with all but one found within 100 au of the largest particle. By 0.35 $\tau_{ff}$ an additional 6 stars have formed, as discussed below.

Columns 13 and 14 of Table~\ref{tab:simlist} give the total number of star particles formed and the number of disks formed for the large mass molecular cloud models. It is clear that turbulence plays an important role in the number of star formed---the more turbulent, the more particles are formed in the aligned cases. In addition, the role of alignment plays some role in the total number formed with 16 stars formed in our fiducial aligned model AT100, 10 stars formed in RT100TH37, and only 3 formed in RT100TH0.  In general the number of star particles formed is in line with those found in \cite{Seifried2013}. For AT100 the largest star formed has a mass of 0.42 \msun, and the average mass is 0.12 \msun. In fact, for all star particles formed among all models, the average mass is around 0.1 \msun.

As shown in Figure~\ref{fig:gevo}, most of the particles have formed in a very small dense region and not throughout the filament. In fact many of these particles form near each other and undergo N-body interactions or form multiple star systems. Figure~\ref{fig:stars} shows the relative distance between each star particle relative to a chosen star particle, noted as the central star. The top row shows these distances for two subsets of particles formed in AT100, with the top left showing those that are close to the central star while the top right shows those that are farther away. The bottom row shows these distances for RT100TH00 and RT100TH37. It is immediately apparent many of the stars formed in AT100 undergo numerous N-body interactions, some of which form bound systems with very small separation distances and short orbit times while others interact only occasionally. However, as shown in the top right panel, not every particle undergoes such interactions.

Since they form fewer stars than AT100, both RT100TH37 and RT100TH00 do not suffer from as many N-body interactions. RT100TH00 never undergoes such an interaction while RT100TH37 forms a single stable binary system.

\subsection{Magnetic Field Structure}
\begin{figure*}
\centering
\setlength{\tabcolsep}{0mm}
\begin{tabular}{ll}
\includegraphics[trim=0.0mm 0.0mm 0.0mm 0.0mm, clip, scale=0.20]{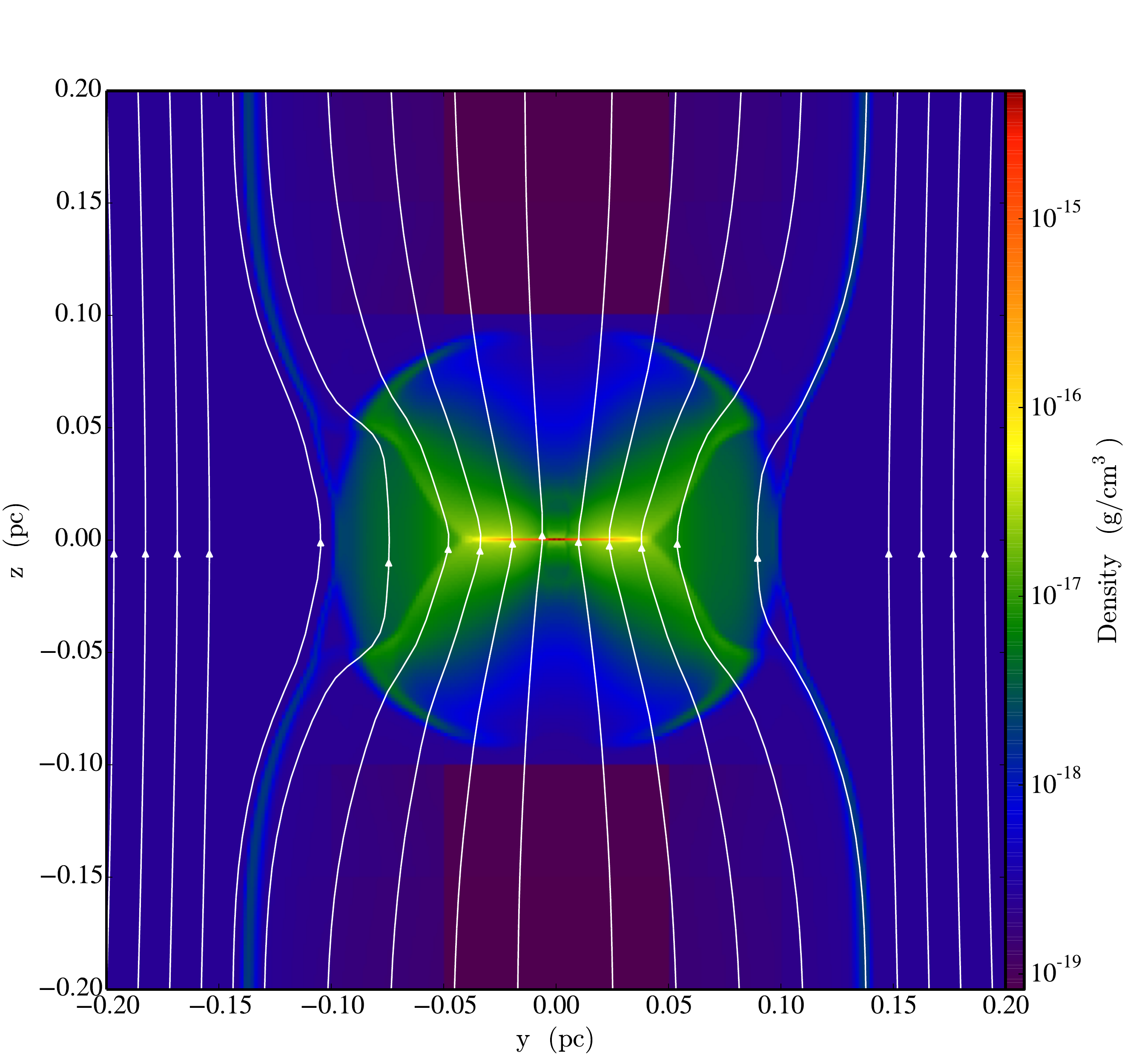}
\includegraphics[trim=0.0mm 0.0mm 0.0mm 0.0mm, clip, scale=0.20]{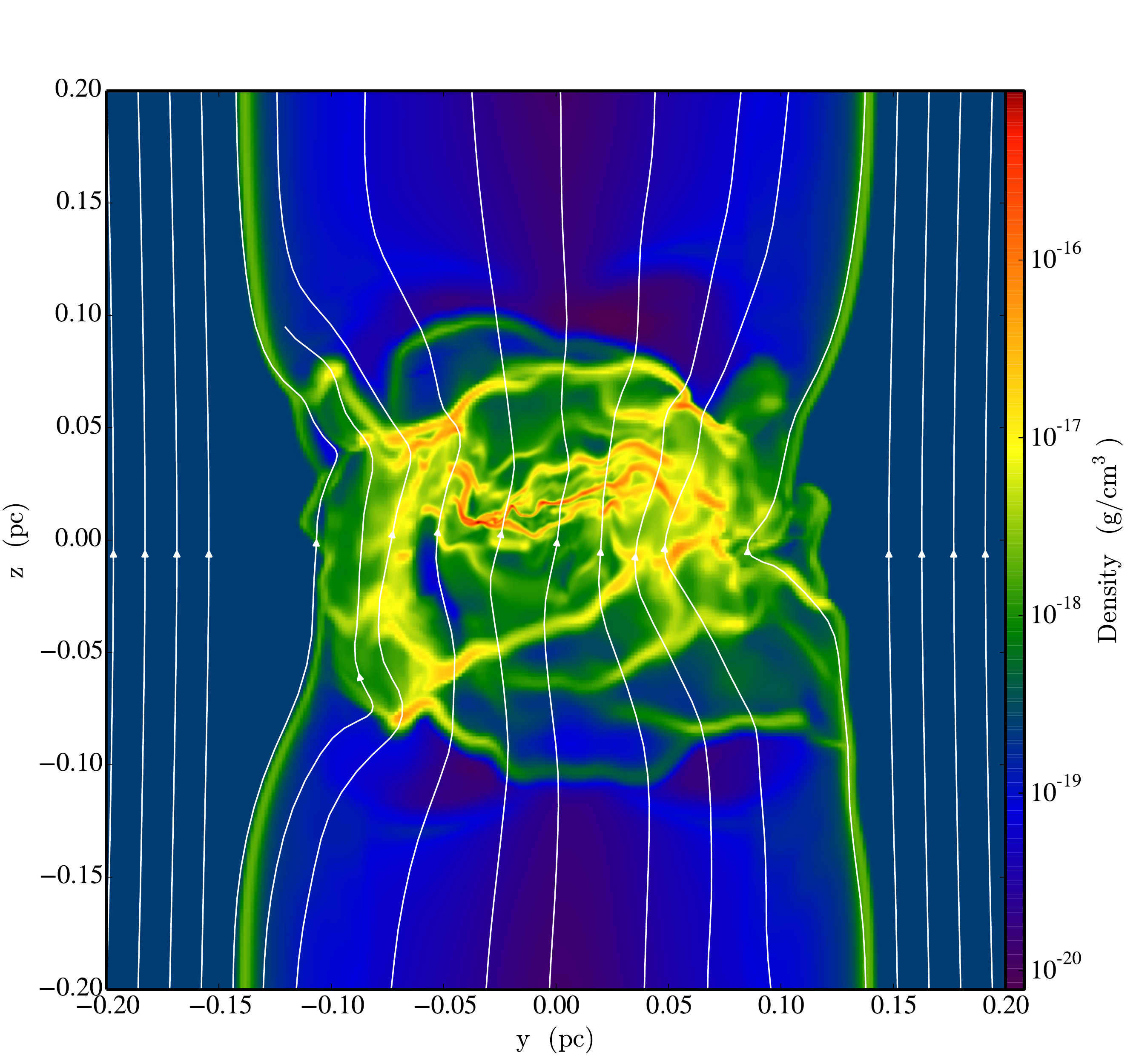}\\
\includegraphics[trim=0.0mm 0.0mm 0.0mm 0.0mm, clip, scale=0.20]{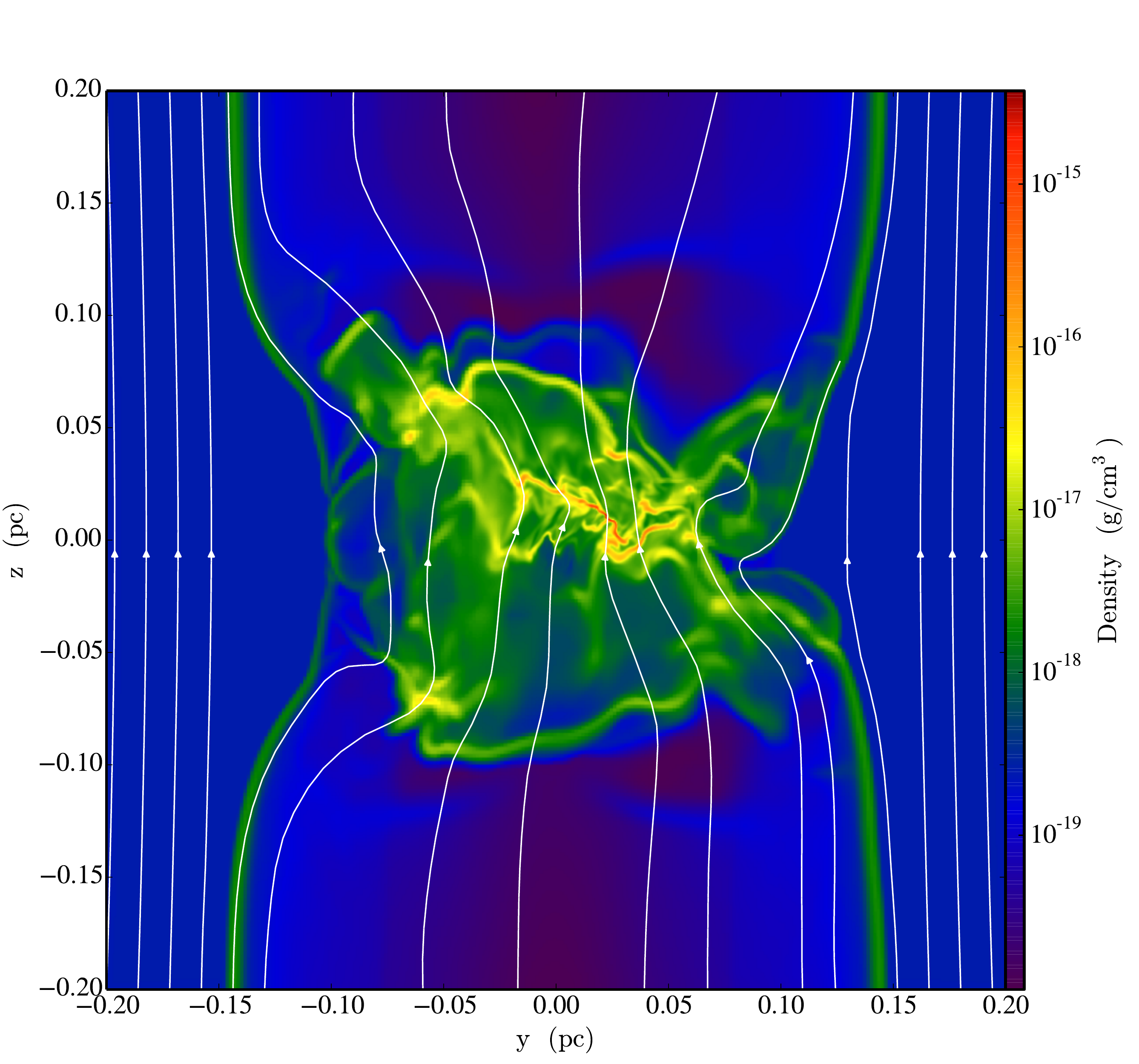}
\includegraphics[trim=0.0mm 0.0mm 0.0mm 0.0mm, clip, scale=0.20]{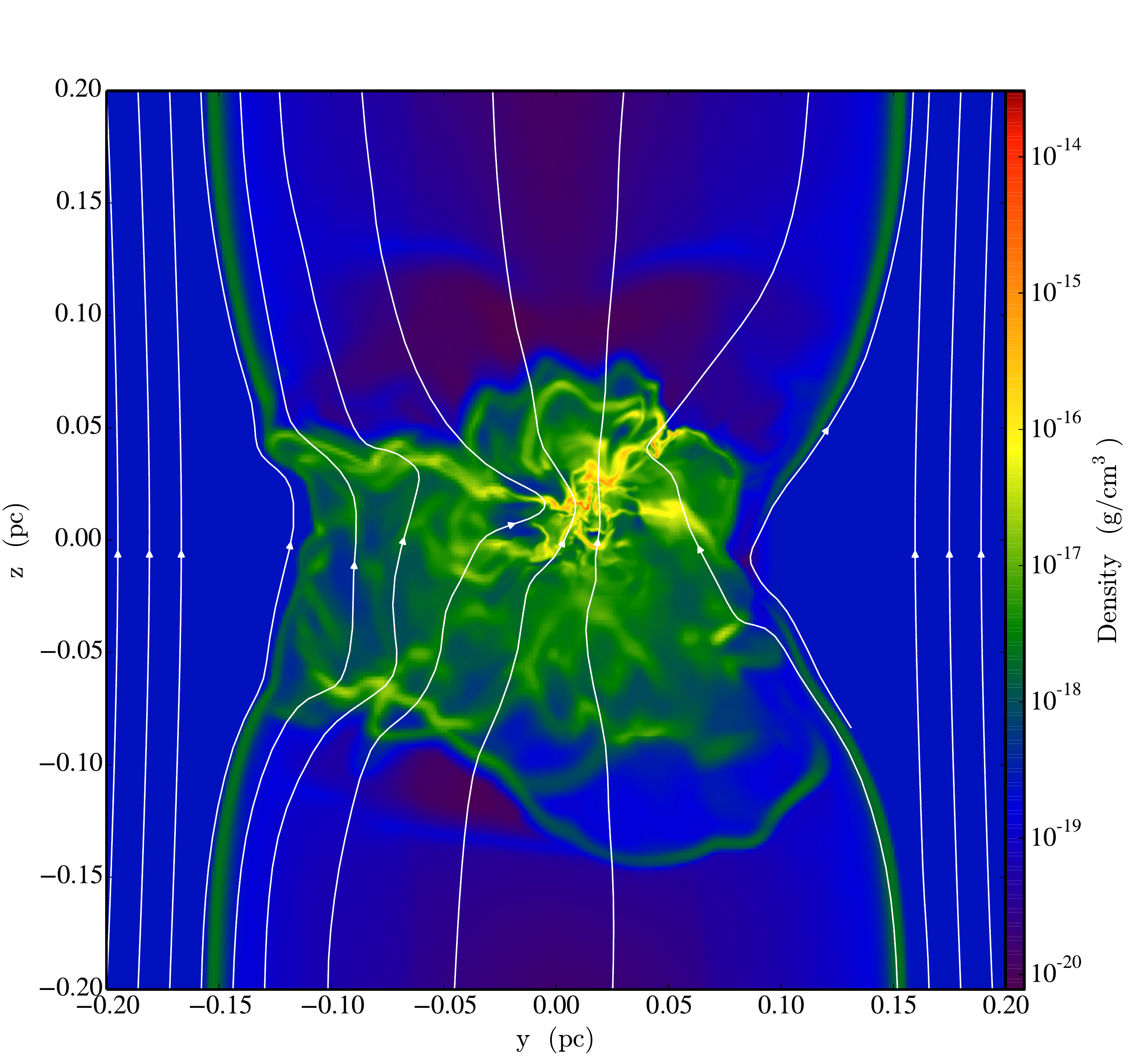}\\
\end{tabular}
\caption{Large scale magnetic fields for {\it Top Left}: AT0, {\it Top Right:} AT100, {\it Bottom Left}: RT100TH0, and {\it Bottom Right:} RT100TH37. The magnetic fields are shown as the white lines overlaid on a density slice through the center of the computational domain. }
\label{fig:LSbfields}
\end{figure*}

Although the magnetic field lines are initially aligned with the $z$-axis, this is not the final arrangement, which is largely dependent on the initial velocity field. Figure~\ref{fig:LSbfields} shows the final magnetic field structure for AT0 (top left panel), AT100 (top right), RT100TH37 (bottom left), and RT100TH0 (bottom right).  At large scales the magnetic field takes on the classic `hourglass'-like profile; this is especially apparent for AT0, which has no turbulent motions. However, this `hourglass' profile is discernible in every model.

This shape is to be expected since the field at cylindrical radii initially outside the cloud is much weaker than the field inside the cloud. As mentioned in \cite{Myers2013}, the field lines have been only slightly bent due to the slightly super-\alfvenic\ turbulence; the amount of bending increases moderately as the turbulent energy increases.

We also note the difference in the large scale density structure between these models. As expected for AT0, most of the mass has gravitationally collapsed along field lines and formed a large, thin disk-like structure. AT100 shows a much more disordered density profile, but retains the property that most of the densest structures are perpendicular to the magnetic field lines. Finally, both RT100TH0 and RT100TH37 have much more misaligned density structures when compared to AT100.

At small scales, particularly at the scale of the star particles, the structure of the magnetic field can be very different. Previous studies that do not include the effects of turbulence \citep[\eg][]{Galli2006,Hennebelle2009,Seifried2011} have shown that the magnetic field lines surrounding the star develop a profile similar to the large scale `hourglass'-like profile. That is, the field is highly compressed through the disk and fans outward. Models that include turbulence \citep[\eg][]{Seifried2013} find that although a recognizable `hourglass' profile forms, it is highly distorted due to turbulence.

\begin{figure*}
\centering
\setlength{\tabcolsep}{0mm}
\begin{tabular}{ll}
\includegraphics[trim=0.0mm 0.0mm 0.0mm 0.0mm, clip, scale=0.20]{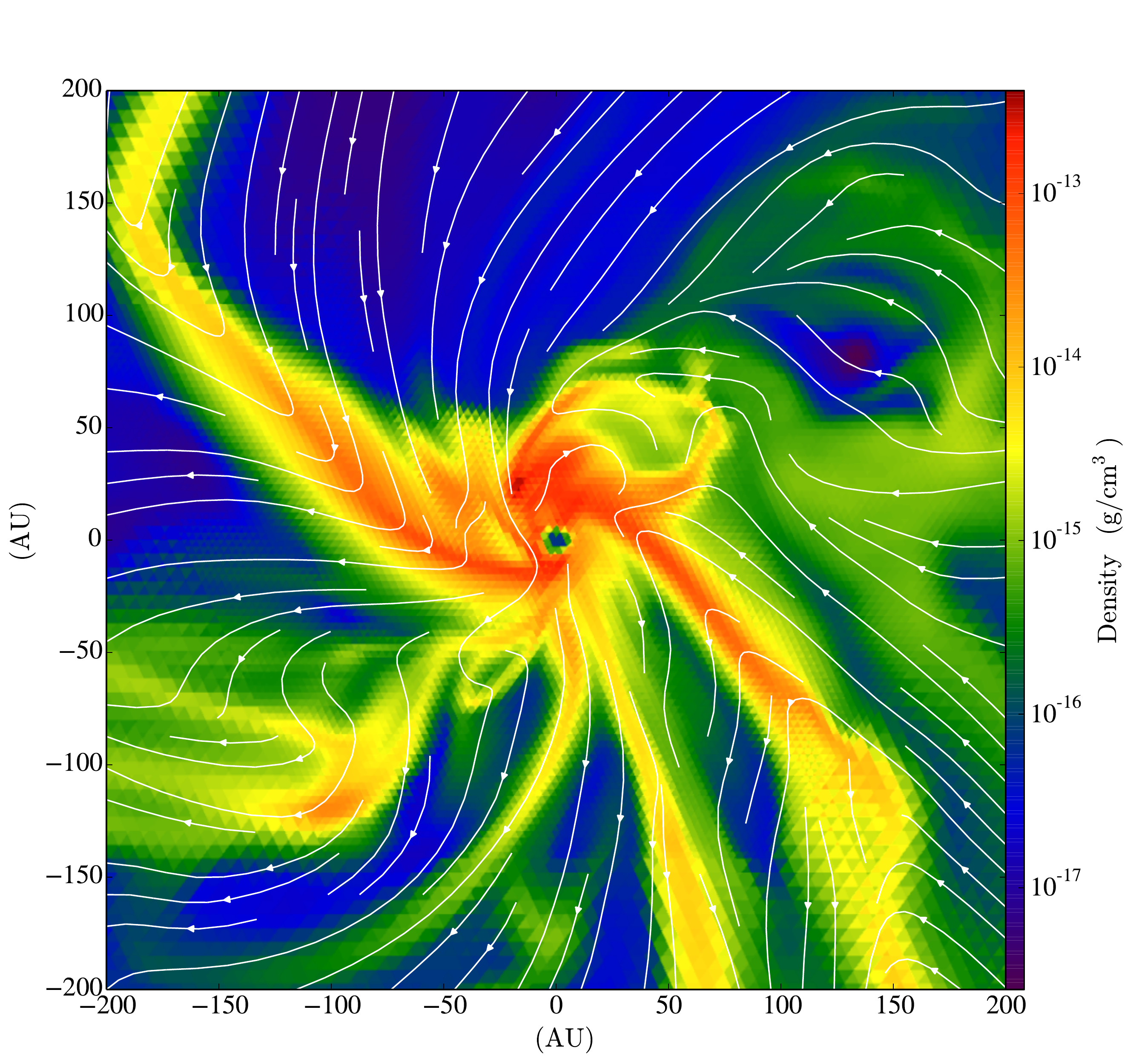}
\includegraphics[trim=0.0mm 0.0mm 0.0mm 0.0mm, clip, scale=0.20]{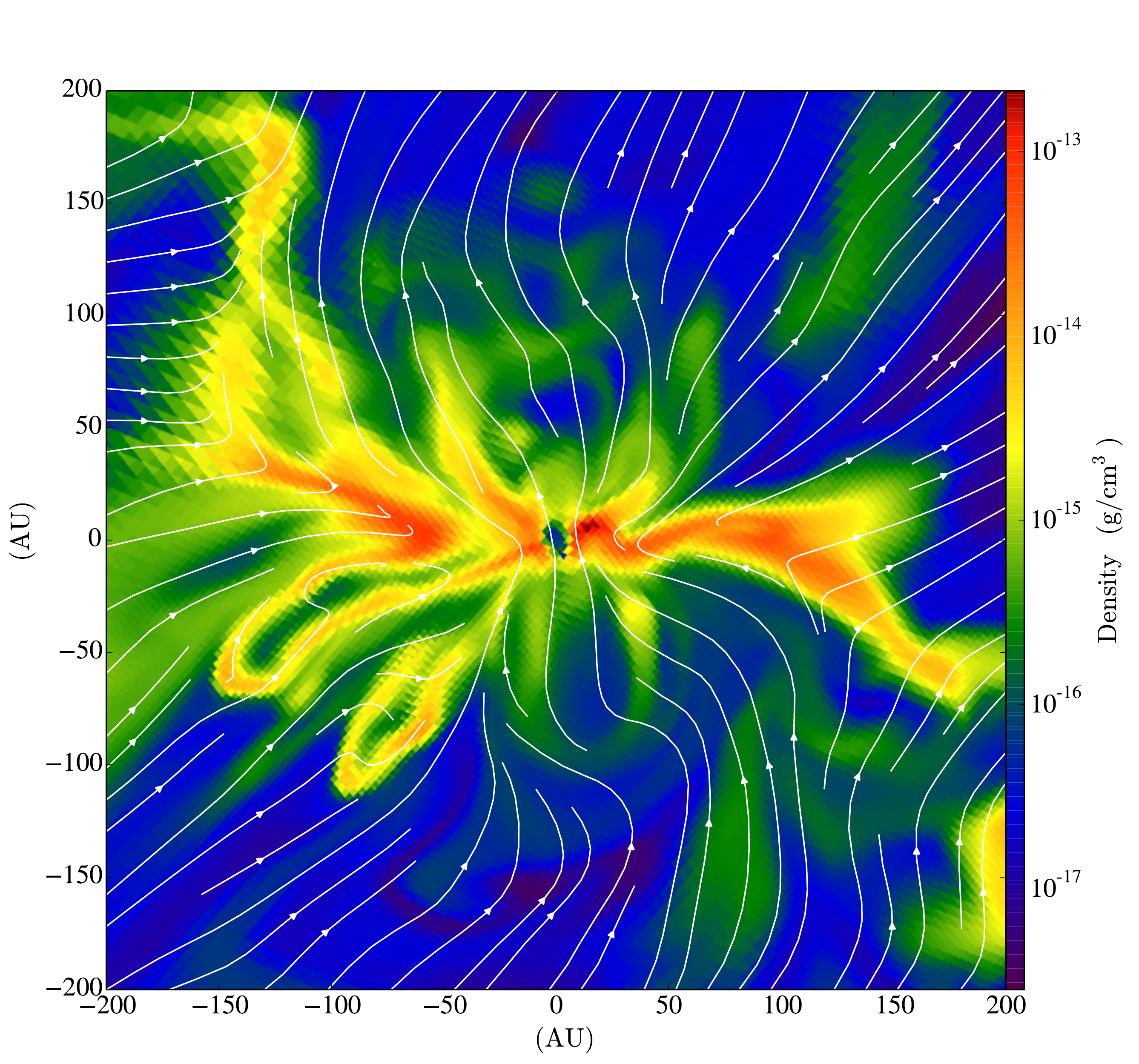}\\
\end{tabular}
\caption{Small scale magnetic fields for the largest particle formed in AT100.  {\it Left Panel:} Face on and {\it Right Panel:} Edge on. The magnetic field is shown as the white lines. The low density zones in the center of each figure represents the accretion zone of the star particle. }
\label{fig:SSbfields}
\end{figure*}

To compare to these studies we have plotted the face-on and edge-on density profiles for the largest star particle formed in AT100 as shown in the left and right panels of Figure~\ref{fig:SSbfields}. We find that the magnetic field does form a disordered `hour-glass' profile in both the edge-on and face-on profiles, which has important consequences on the braking efficiency within the disk.

The classic magnetic braking mechanism is dependent on the formation of a toroidal magnetic field, built up by the orderly large-scale rotational motion of the accreting gas \citep[\eg][]{Mouschovias1980}. In idealized non-turbulent cases, the magnetic field structure fans outward from a nominal disk radius $R_0$ to regions far removed from the disk, $R$. This significantly increases the magnetic braking efficiency by a factor of $(R/R_0)^4$ \citep{Mouschovias1985,Joos2012}. However, in models with sufficiently  random turbulence the build up of a strong toroidal magnetic field is prevented. As pointed out by \cite{Seifried2013}, large scale rotation is disrupted, which lowers the efficiency of magnetic braking by preventing a strong coupling between the inner parts close to the disk to the outer regions of the cloud. It should also be noted that simply enforcing a misalignment between the angular momentum vector and the magnetic field also decreases the magnetic braking \citep{Hennebelle2009,Ciardi2010,Joos2012}.

\begin{figure}
\centering
\setlength{\tabcolsep}{0mm}
\includegraphics[trim=0.0mm 0.0mm 0.0mm 0.0mm, clip, scale=0.450]{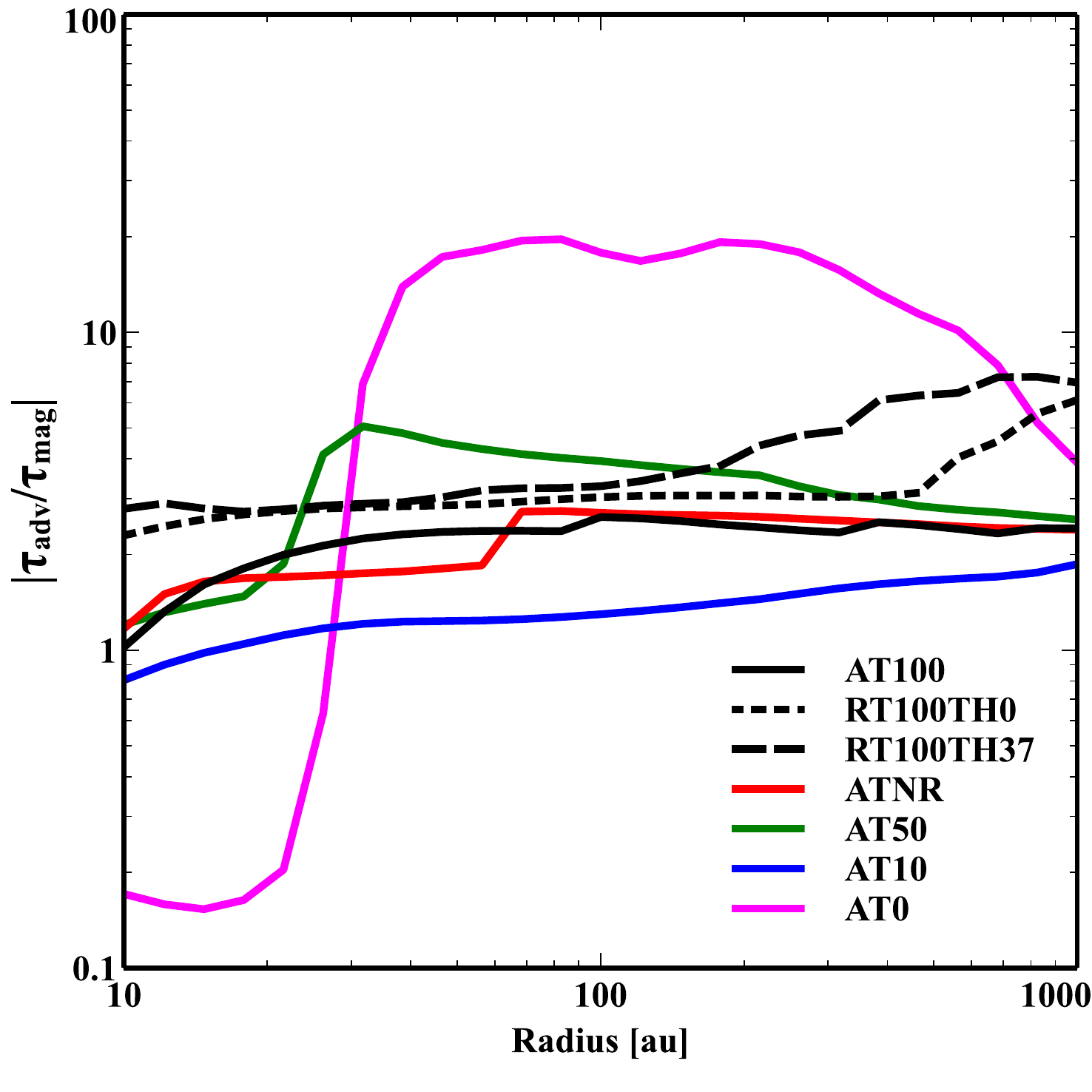}
\caption{ Ratio of $\tau_{\rm adv}$ to $\tau_{\rm mag}$ as a function of radius for models with differing initial turbulent kinetic energies. AT100 is shown as the solid black line, RT100TH37 is the dashed black line, and RT100TH0 is the dotted black line, ATNR is the red line, AT50 is the green line, AT10 is the blue line, and AT0 is the magenta line. }
\label{fig:taus}
\end{figure}

For a quantitative insight into the magnetic braking processes, we  follow \citet{Seifried2011,Joos2012,Li2013} and consider the equation governing the angular momentum of the accreting gas \citep{Kulsrud2005}:
\be
\frac{d}{dt}\int {\bf r\times}\rho{\bf  v} dV=\int{\bf r\times}\ppbyp{\rho{\bf v}}{t}dV.
\ee
The fluid equation of motion gives $\partial\rho{\bf v}/\partial t$. The advective term in this equation, $-\nabla\cdot\rho{\bf vv}$, leads to a surface integral that gives the flow of angular momentum into the volume \citep{Kulsrud2005}; this is termed the advective torque:
\be
{\bf \tau}_{\rm adv}=-\int ({\bf r\times}\rho{\bf v}){\bf v\cdot dS},
\ee
or converting to a volume integral
\be
{\bf \tau}_{\rm adv}=-\int ( v\cdot \nabla ( {\bf r \times}{\rho {\bf v}} ) + \rho({\bf r \times v}) \nabla \cdot {\bf v} ) dV.
\ee
The magnetic force in the equation of motion leads to the magnetic torque

\be
{\bf \tau}_{\rm mag} = \frac{1}{4\pi} \int dV [ {\bf r} \times (( \nabla \times {\bf B} ) \times {\bf B} ) ].
\ee
These integrals are evaluated over a disk with a height of 50 au and a radius that varied between 10-1000 au. Figure~\ref{fig:taus}  shows the absolute value of the ratio between these two torques, $|\tau_{\rm adv}/\tau_{\rm mag}|$, for models over a range of initial turbulent kinetic energies for the largest particle formed in each simulation (see \S\ref{sec:Eturb} below). If no particle forms, as in the case of AT0, we calculate this ratio around the center of mass of the largest clump formed. We find that for all of our aligned models, this ratio stays very near unity between 1 au and 100 au. This means that whatever angular momentum is gained by the gas as it collapses near the star, is very nearly completely lost by the magnetic field. Only at slightly larger radii, $>$100 au, does the advective torque become greater than the magnetic torque. This suggests that even though the gas is turbulent, the magnetic field remains strong enough to prevent the collapsing gas from keeping its angular momentum. The situation is worse for AT0. Here the magnetic torque is very efficient at removing angular momentum from the gas, keeping the ratio below one. Only for the initially random turbulent models, RT100TH0 and RT100TH37, does this ratio stay appreciably above unity over the entire range of radii.

\subsection{Effect of Turbulent Alignment}

We begin by looking at the differences between AT100, RT100TH0, and RT100TH37, which span the range of possible initial misalignments in our simulations. The turbulence in AT100 is initially aligned with the magnetic field using the procedure outlined in Appendix.~\ref{turbalign}. RT100TH0 uses the same procedure used for AT100 with the exception that a single shell is used instead of a series of shells. This creates a misalignment profile that guarantees alignment for the cloud as a whole, but not for the interior of the cloud. RT100TH37 uses a completely random initial turbulence field and, consequently, a large initial misalignment.

The first difference between these models is the time at which the first star particle forms. For AT100 the first particle forms by 0.21\tffm\ and for RT100TH0 it forms by 0.22\tffm, whereas it takes until 0.34\tffm\ for RT100TH37. Since this is beyond the nominal simulation run time, we have allowed all three models to run to nearly $0.4\tff$ to ensure a proper comparison.

We find that, although a total of 16 star particles form in the AT100 model, {\it no disks, as defined in Section \ref{sec:crit} above, are formed around any particle at any time in this model}. In fact, as we will discuss below, {\it no disks are formed in any model where the turbulence is initially aligned with the magnetic field with an initial mass-to-flux ratio of 2}. RT100TH37, on the other hand, forms ten star particles, of which at least one has a prominent disk. Finally, RT100TH0 forms three star particles, one of which has a disk.  It is striking that the run with no initial turbulent misalignment formed no disks (AT100), whereas the runs that included
turbulent misalignment, even if the turbulence was globally aligned, did (RT100TH37 and RT100TH0).

\cite{Seifried2013} found that the number of disks formed is strongly dependent on the turbulence. They presented three identical models where they vary only the turbulent seed and found that on average one disk forms for every five to nine particles formed. This is consistent with the results from RT100TH37 and RT100TH0 and suggests that turbulent misalignment plays an important role in the formation of disks.

\begin{figure*}
\centering
\includegraphics[trim=0.0mm 0.0mm 0.0mm 0.0mm, clip, scale=0.55]{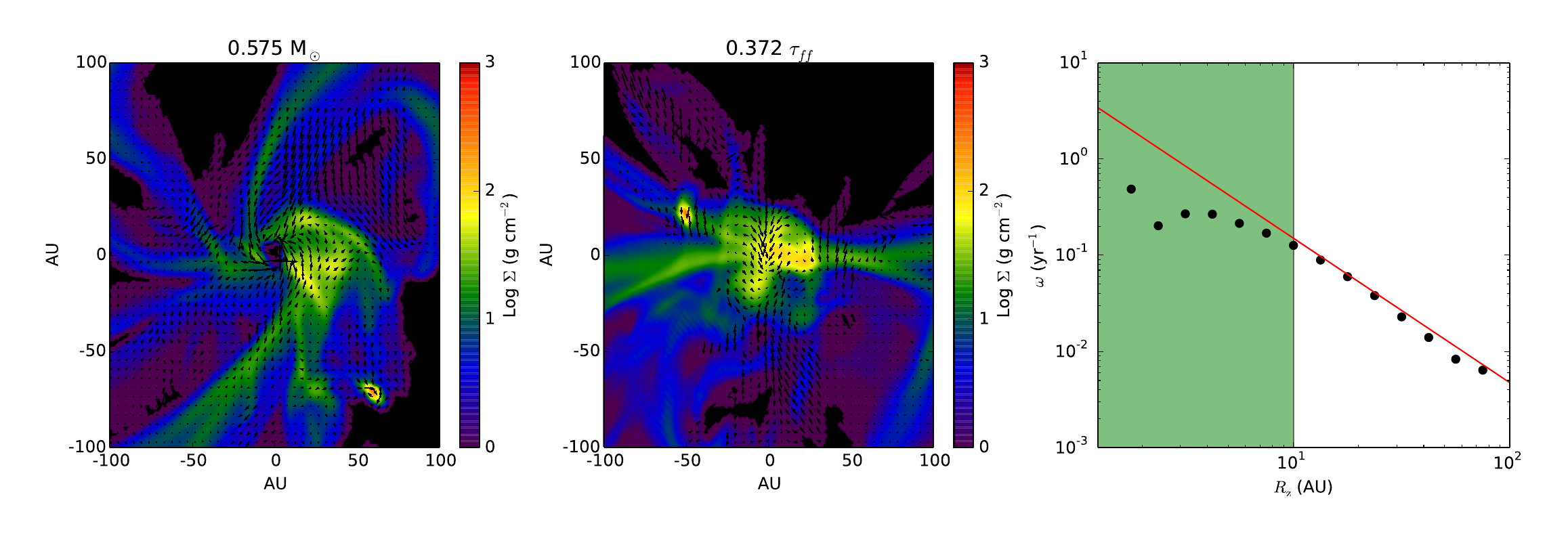}
\includegraphics[trim=0.0mm 0.0mm 0.0mm 0.0mm, clip, scale=0.55]{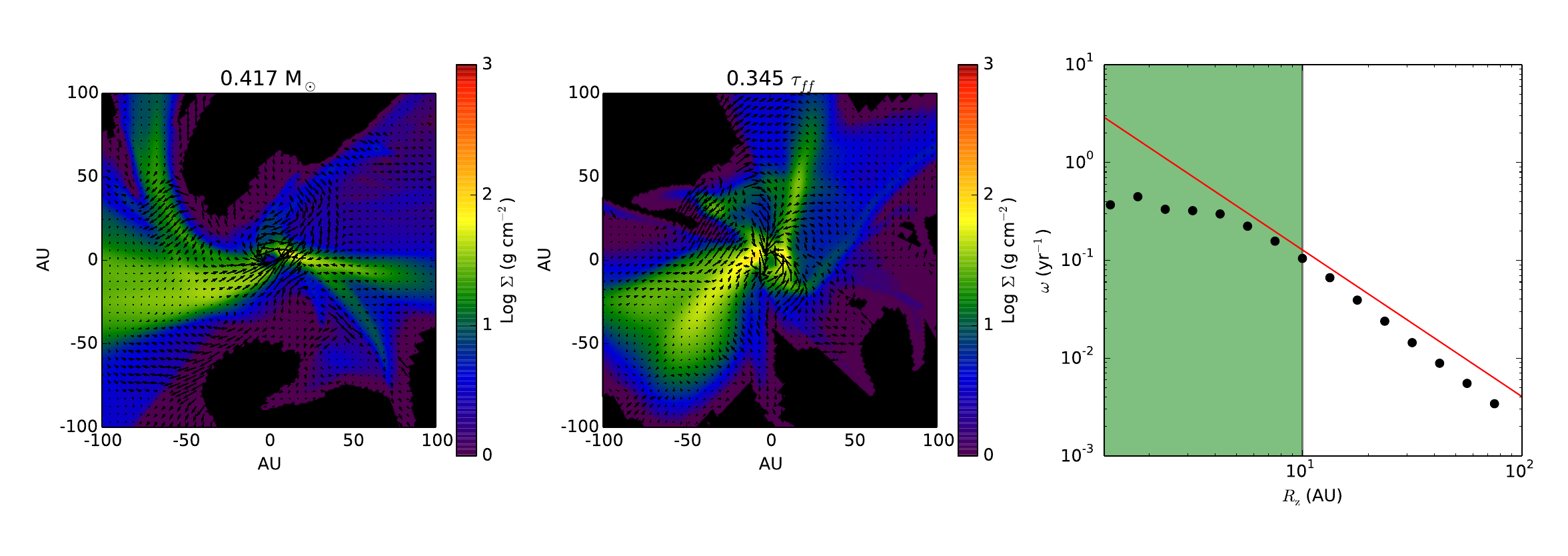}
\includegraphics[trim=0.0mm 0.0mm 0.0mm 0.0mm, clip, scale=0.55]{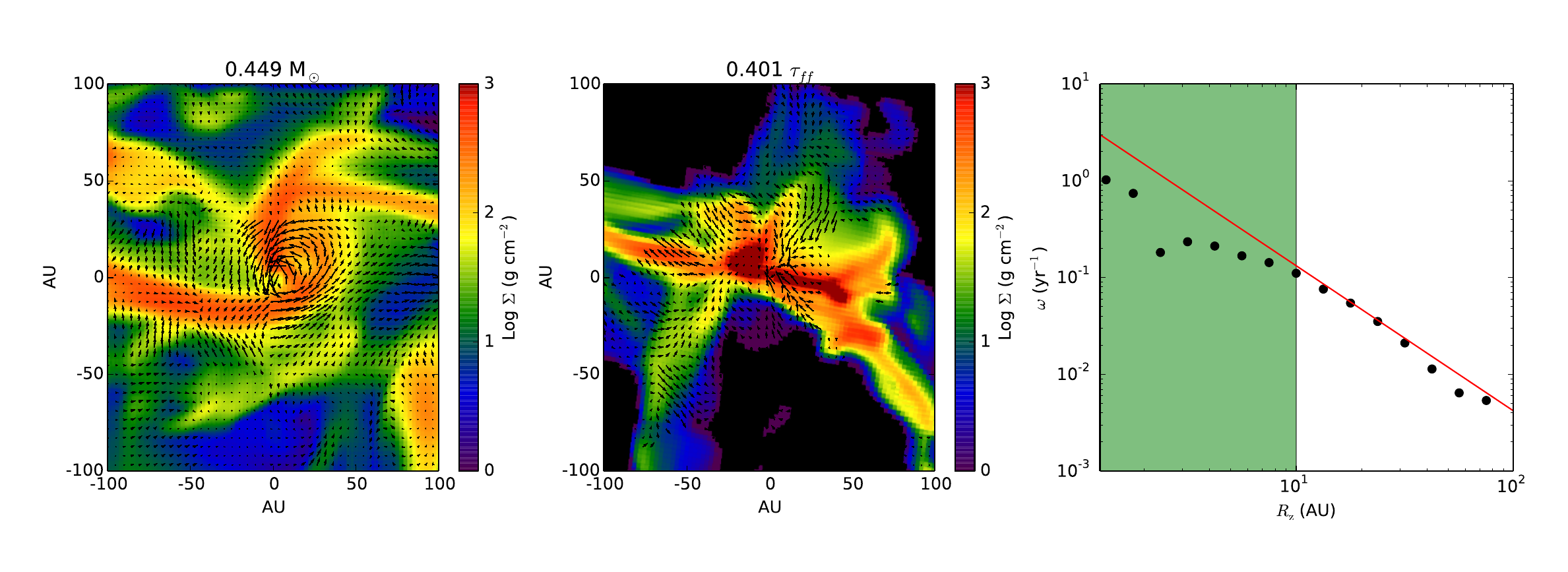}
\caption{ Density projections for the largest star formed in {\it Top Panel:} AT100, {\it Middle Panel:} RT100TH0, and {\it Bottom Panel:} RT100TH37. The left column shows the face-on density projection while the middle column shows the edge-on density projection. The black arrows represent the velocity vectors. The right column shows the velocity profile for the disk. The red (solid) line is the analytic Keplerian angular velocity using the stellar mass. The black points represent the mean angular velocity as a function of the cylindrical radius $R_{z}$. The green region represents regions within the star particle's accretion zone and shows where the accretion zone alters the gas properties.
The left and middle columns include only gas denser than $10^9$ cm$^{-3}$ (Criterion 5).  The final stellar mass is given above the face-on images for each model. Note that disks are formed in both RT100TH0 and RT100TH37 but not in AT100.}
\label{fig:alignp}
\end{figure*}

The bottom panel of Figure~\ref{fig:alignp} shows the face-on, edge-on, and Keplerian profile for the disk formed in RT100TH37. In both the face-on and edge-on profiles, the disk-like structure is plainly visible and satisfies all criteria up to a radius of 50 au. The middle panel shows the disk formed in RT100TH0, which also satisfies the criteria up a radius of 50 au. More quantitatively, the disk system formed in RT100TH37 has a stellar mass of M$_*$=0.45 M$_\odot$, disk mass of M$_d$=0.36 M$_\odot$, and a radius of 55 au. The disk system formed in RT100TH0 has a stellar mass of M$_*$=0.42 M$_\odot$, disk mass of M$_d$=0.02 M$_\odot$, and a radius of 91 au. As can be seen in Figure~\ref{fig:alignp}, the disk in RT100TH0 is much denser and more compact than the one formed in RT100T37.

The top panel shows the profiles for the largest star formed in AT100. Although there is some evidence for Keplerian rotation, both density projections show little evidence for a coherent disk. In fact, the gas surrounding this particle is never rotationally supported.

\begin{figure*}
\centering
\includegraphics[trim=0.0mm 0.0mm 0.0mm 0.0mm, clip, scale=0.70]{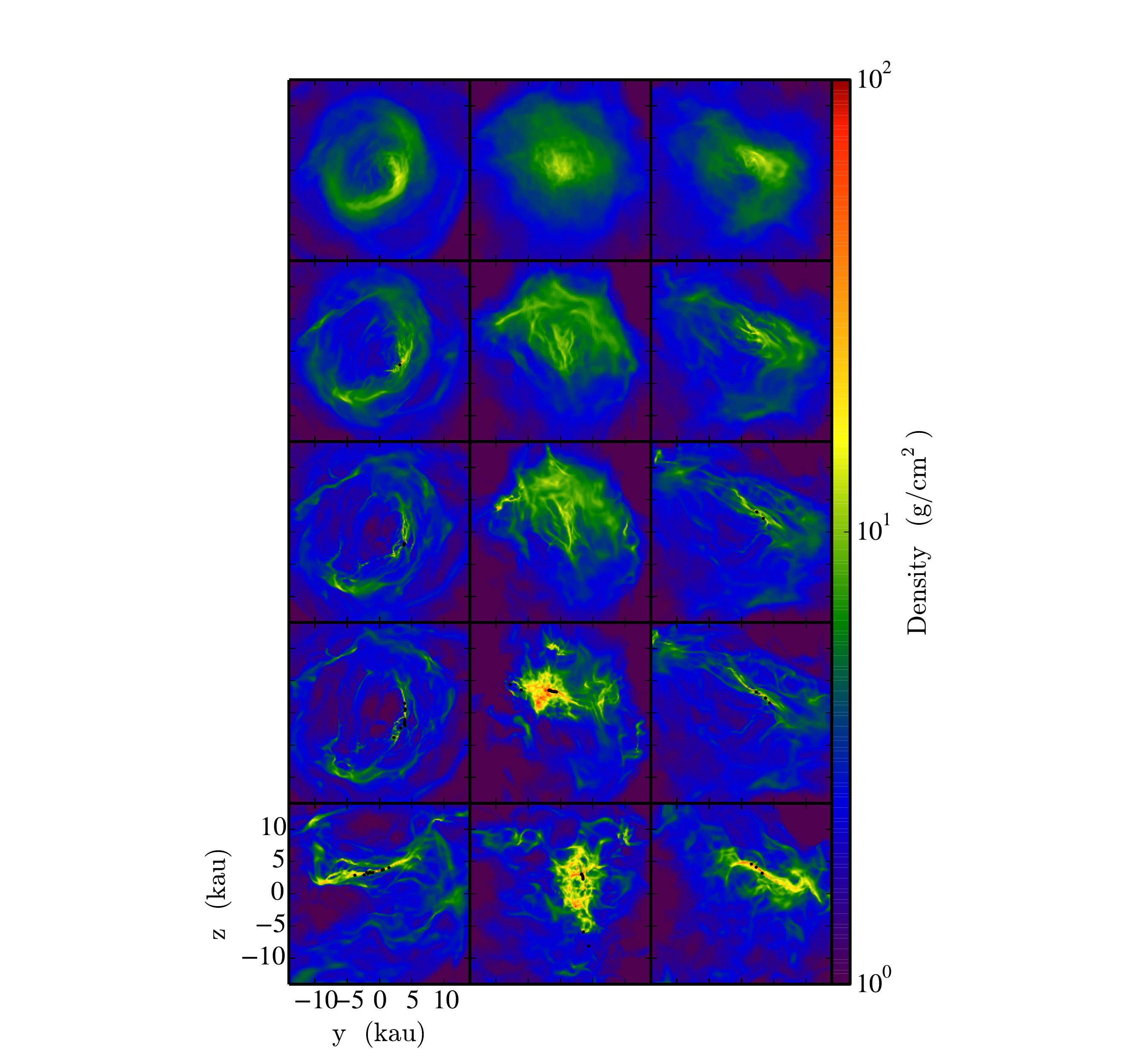}
\caption{Density projections showing the large scale evolution of AT100, RT100TH37, and RT100TH0. The density is projected along the $z$-axis. AT100 is shown in the first column, RT100TH37 in the second, and RT100TH0 in the third. Time runs forward starting at $t=0.1\tff$ for the first row, $t=0.2\tff$ in the second row, $t=0.3\tff$ in the third row, and $t\simeq0.4\tff$ in the fourth row. The last row shows the same time but for a projection taken along the $x$-axis. Black points show the positions of the star particles. }
\label{fig:alignls}
\end{figure*}

Figure~\ref{fig:alignls} shows the large scale evolution of AT100 (first column), RT100TH37 (second column), and RT100TH0 (third column). Each rows shows a density projection along the $z$-axis except for the last row which shows a projection along the $x$-axis. The final density structure between each of these models is very different. As mentioned above, most of the gas in AT100 forms a long arcing filamentary structure. A similar structure is seen RT100TH0 although the filament is more misaligned from the initial magnetic field, as seen in the bottom right panel. RT100TH37 shows the most disordered and shows little evidence for any filamentary structure. Observationally, the filamentary structure seen in AT100 is more natural than the structure found in RT100. This may be a product of our simplified initial condition of a spherical cloud rather than a more complex physical shape.

\subsection{Effect of Varying Turbulent Energy}
\label{sec:Eturb}

We next turn our attention to the effect of varying the amount of turbulence present in the initial cloud on the formation of disks. Four initial turbulent energies are studied, ranging from \alphavirm = 0.0,0.1,0.5, and 2.0.  These correspond to a range of \alfvenic\ Mach numbers between sub-\alfvenic ($\ma = 0.3$) to moderately super-\alfvenic ($\ma = 1.6$). In all models the turbulence is thermally supersonic, with Mach numbers ranging between 2.8 and 15.

\begin{figure*}
\centering
\includegraphics[trim=0.0mm 0.0mm 0.0mm 0.0mm, clip, scale=0.70]{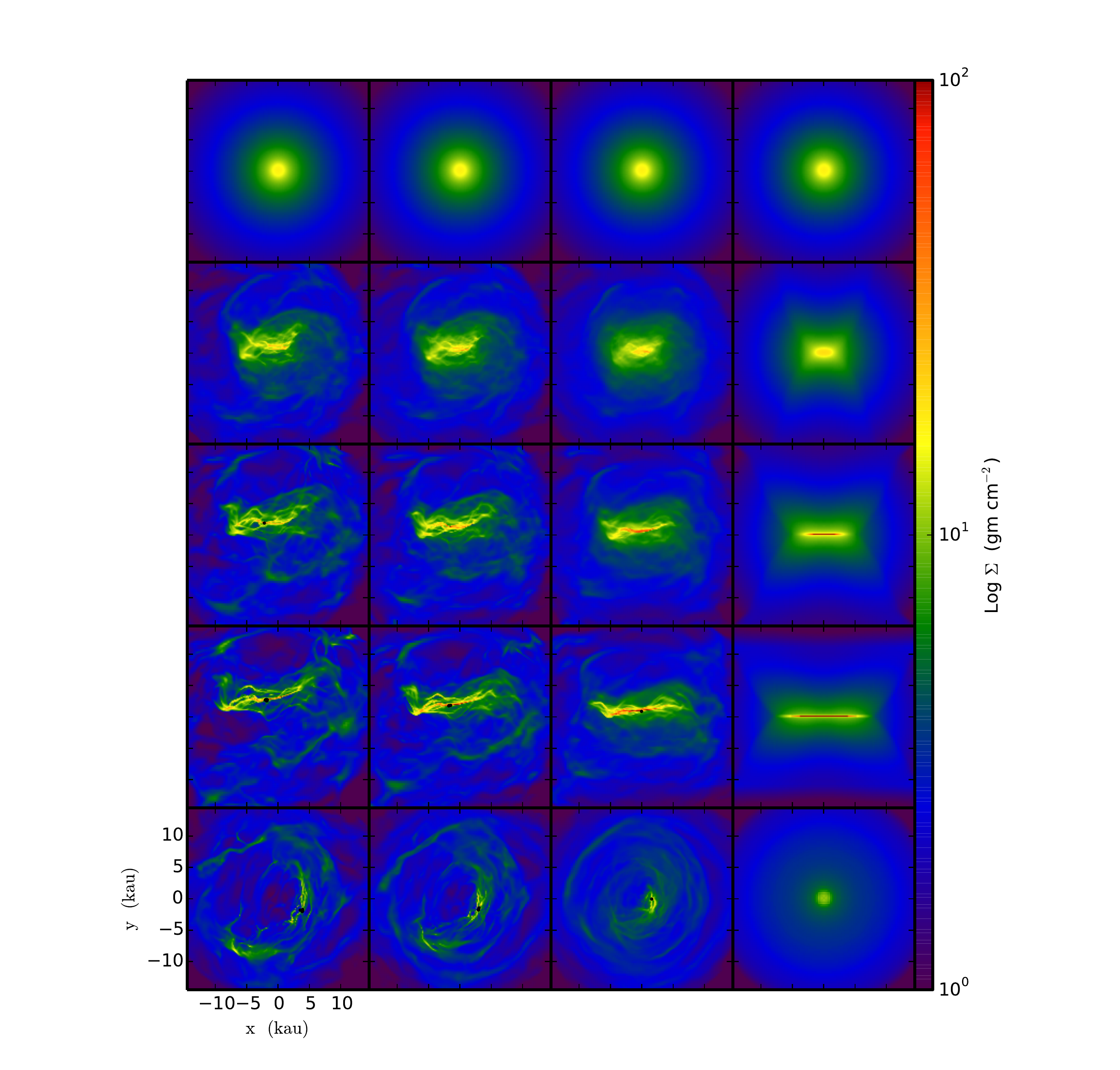}
\caption{Column density projections at four times for {\it Left}: ATNR, {\it Left Middle}: AT50, {\it Right Middle}: AT10, and {\it Right}: AT0. Projections are taken along the $x$-axis (in units of 1000 au) except for the last row, which shows a projection along the $z$-axis at the final time. The viewing area is set to include the central 15000 au of the core, showing the large scale evolution of the cloud. The black points show the location of the star particles.}
\label{fig:turbpanel}
\end{figure*}

Figure~\ref{fig:turbpanel} shows the global evolution of models ATNR, AT50, AT10, and AT0. The first four rows show the density projections along the $x$-axis at $t/\tff$ = 0.0, 0.1, 0.2, and 0.3 respectively. The last row shows the density projection along the $z$-axis at $t/\tff=0.3$. The shape, extent, and orientation of the filament, as seen in the last row of Figure~\ref{fig:turbpanel} for ATNR, AT50, and AT10, remains largely the same between each run, except for AT0, which instead forms a single large disk-like structure. Here we define the filament as the large dense structure formed near the mid-plane of the $x$-axis projections and off-center in the $z$-axis projection in Figure~\ref{fig:turbpanel}. However, the size and compactness of the filament depend on the initial Mach number, with smaller widths and sharper features forming with higher Mach numbers.

There is a large spread in the number of star particles formed in each simulation. Quantitatively, ATNR forms ten particles, AT50 forms four, AT10 forms only one, and AT0 does not form any particles.  The result for AT0 is due to both the pressure support from the barotropic equation of state and the strong magnetic field. However, it is likely that a particle would eventually form in the center of the pseudo-disk if the simulation were run longer. Since the underlying driving pattern is the same for each model, this difference is attributed solely to the sonic Mach number. This result compares well with recent simulations of turbulent magnetized clouds by \cite{Federrath2012}, which showed that the specific star formation rate generally increase with the sonic mach number.

However, {\it no evidence is found for a Keplerian disk in any run with fully aligned turbulence;} that is, at no time does the gas around any of the sink particles in these simulations fulfill all the disk criteria--in particular, the requirement that the gas is rotationally supported routinely fails.

\section{Evolution of \muphim}
\label{sec:comp}

The normalized mass-to-flux ratio, \muphim, determines whether a rotationally supported disk can form since it is proportional to the ratio of the magnetic braking time, $t_{\rm mb}$, to the free-fall time, $\tff$ \citep{Mestel1984}. One can estimate the magnetic braking timescale as the time it takes for a \alfven\ wave to redistribute angular momentum from the inner part of the collapsing core into the outer part or into the ambient medium, $t_{\rm mb} \sim  Z/v_{\rm A}$, where $Z$ is the distance an \alfven\ wave must travel to sweep up gas with a moment of inertia equal to that of inner part of the collapsing core and $v_{\rm A}=B/\sqrt{4\pi\rho}$ is the \alfven\ speed in the outer part of the core or in the ambient medium.

It follows that
\be
\frac{t_{\rm mb}}{\tff}\propto \frac{Z\rho^{1/2}}{B}\cdot (G\rho)^{1/2}\sim G^{1/2} \frac{\rho ZR^2}{B R^2}\sim \muphi.
\ee

For the particular case in which the \alfven\ waves radiate outwards approximately spherically from the collapsing core, rather than being confined to a flux tube with a much smaller opening angle, the analysis given by \citet{Mestel1984} leads to
\be
\frac{t_{\rm mb}}{t_{\rm ff}}= 0.26\left(\frac{\rho_c}{\rho_0}\right)^{1/10} \muphi
\ee
\citep{McKee1993},  where the difference in the coefficient from that reference is because they used $c_{\Phi}=0.126$ (see equation \ref{eqn:cmass}) and where $\rho_c$ and $\rho_0$ are the densities of the collapsing core and the surrounding material, respectively. An analogous result has been given by \citet{Joos2012}, who considered a disk geometry and compared the braking time to the rotation period instead of the free-fall time.

This expression is modified after a protostar forms in the core. The magnetic braking time is unchanged, but (assuming spherical symmetry) the density that enters the free-fall time is the mean density of all the matter in the collapsing core, including that in the star. As a result, the free fall time is reduced by a factor $(\rho/\rho_\tot)^{1/2}=(M/M_\tot)^{1/2}$, where $M_\tot=M+M_*$ is the mass of both the gas and the star, and the effective value of \muphim\ that measures magnetic braking becomes
\be
\mu_{\Phi,\,\rm eff}=2\pi\sqrt G\left(\frac{\sqrt{MM_\tot}}{\Phi}\right).
\label{eqn:muphieff}
\ee

\begin{figure}
\centering
\includegraphics[trim=10.0mm 0.0mm 0.0mm 0.0mm, clip, width=0.85\columnwidth]{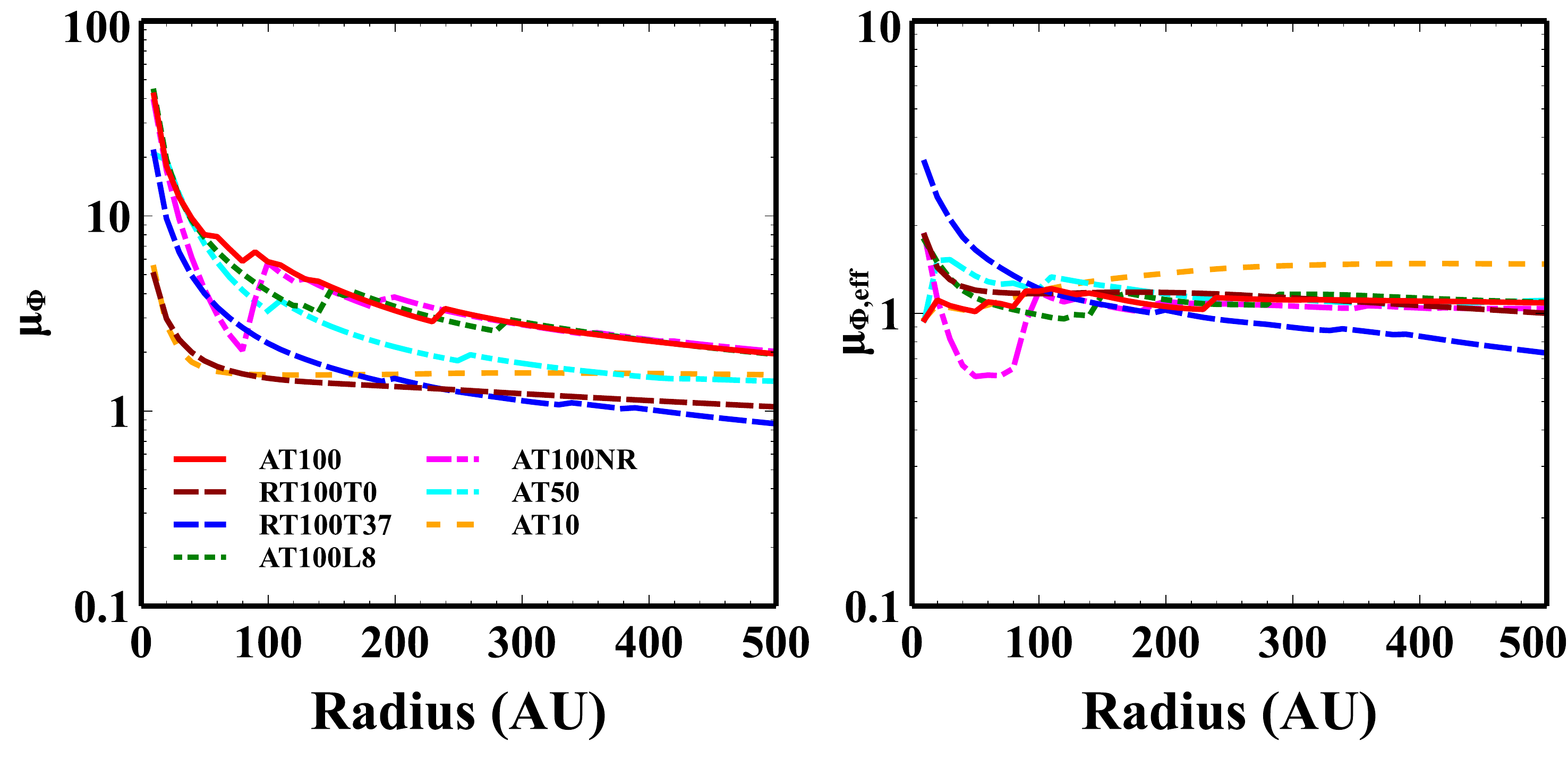}
\caption{ {\it Left Panel}: mass-to-flux ratios using the total mass versus radius for the largest particles formed. {\it Right Panel}: mass-to-flux ratios (Eqn \ref{eqn:muphieff}). The legend gives the line style and line color for each model. }
\label{fig:mvr}
\end{figure}

The left panel of Figure~\ref{fig:mvr} shows $\muphi(M_\tot)$--i.e., the mass-to-flux ratio based on the total mass--as a function of radius for the largest particle formed in AT100, RT100TH37, RT100TH0, AT100L8, and all models where the initial turbulent energy is varied; this excludes AT0 as no particle is formed.  For every model shown \muphim\ decreases with distance to the particle. At small radii \muphim\ becomes very large, $>10$, but it returns to the initial value by $\sim100-200$ au in all models. The rise in \muphim\ near the star is largely due to fact that the flux associated with the gas that was accreted onto the star has been left behind.

The right panel of Figure~\ref{fig:mvr} shows $\mu_{\Phi,\,\rm eff}$, the effective mass-to-flux ratio defined above. Here we find that even at small radii, the effective mass-to-flux ratio remains nearly constant at about the critical value for most of our models. In fact, only RT100TH37 shows an appreciable increase at small radii, and RT100TH0 is the only other run that shows a monotonic increase of $\mu_{\Phi,\,\rm eff}$ as $r$ decreases. It is significant that the only two cases that show Keplerian disks have effective mass-to-flux ratios that monotonically
increase inward.

\begin{figure}
\centering
\includegraphics[trim=0.0mm 0.0mm 0.0mm 0.0mm, clip, width=0.85\columnwidth]{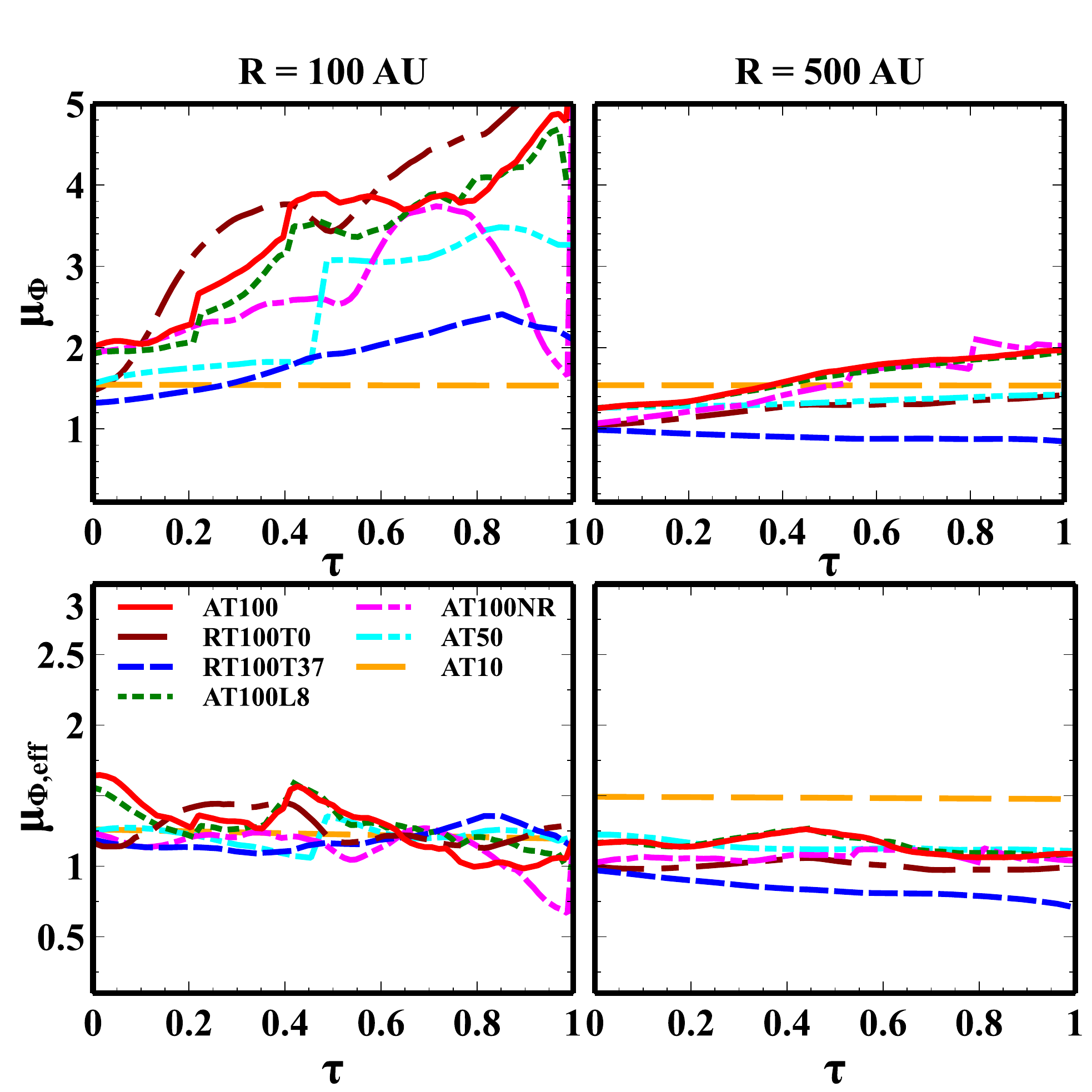}
\caption{ Computed mass-to-flux ratios versus particle lifetime for each model. The legend gives the line style and line color for each model. The top row computes \muphim\ using the total mass where the bottom row uses equation (\ref{eqn:muphieff}) for the critical mass. Each column computes \muphim\ at the radius given in the title. }
\label{fig:mvt}
\end{figure}

Finally, we show the temporal evolution of \muphim\ and $\muphieff$ over the lifetime of the largest particle formed in each simulation in Figure~\ref{fig:mvt}. The particle lifetime is defined as beginning when the particle first forms and stops when the simulation is ended. The top row computed \muphim\ using the total mass whereas the bottom row uses  equation (\ref{eqn:muphieff}). We calculate \muphim\ at 100 and 500 au shown in the left and right columns respectively.  At large radii, 500 au, the particle mass is small compared to the enclosed gas mass and the right columns show that \muphim\ varies little, remaining nearly constant and near the initial value of $\muphi=2$, over the lifetime of the particle. Only at 100 au is there an increase in \muphim\ with particle lifetime (top left panel), and this occurs only in models with a large initial turbulent kinetic energies, \eg AT100 and RT100TH37.  However, much of this is accounted for by the accretion onto the star particle and the subsequent loss of magnetic flux. The bottom panel of Figure.~\ref{fig:mvt} shows the same temporal evolution but for $\muphieff$. We find that the mass-to-flux ratio does not vary much over the lifetime of the star particle even at 100 au; the increase in $\muphieff$ for the non-aligned cases noted above occurs at smaller radii.

Finally, we note that simple numerical dissipation due to the discretization of the computational grid also contributes slightly to the loss of magnetic flux. However, since we find little difference between AT100 and AT100L8, numerical dissipation is not likely to be a dominant source of changes in magnetic flux in our models. However, it is obvious that initial misalignment due to random turbulent motions are necessary for the formation of protostellar disks.

\section{Comparison with Previous Studies}
\label{sec:prevstudies}

We have shown here that rotationally supported disks can form in gravitationally contracting molecular clouds when the initial mass-to-flux matches observed values \cite[\eg][]{Troland2008} only when the initial turbulence is random. This suggests that turbulence plays an essential, if not primary, role in the formation of protostellar disks. Our results are thus consistent with a number of recent studies that have found that turbulent motions in molecular clouds promote the formation of rotationally supported protostellar disks \citep[\eg][]{Joos2013, Seifried2013, SantosLima2012}, although these studies differ on the mechanism that promotes disk formation. \cite{Seifried2013} performed a series of MHD simulations that spanned a wide range of cloud masses, from 2.6 \msun to 1000 \msun, turbulent kinetic energies, and mass-to-flux ratios. Specifically, they ran a 300 \msun\ model that is based on \cite{Myers2013}, which provides an ideal comparison to the models we have run here. Although there are many differences between their model and the models here, specifically the magnetic field profile, the turbulent velocity profile and the equation of state used, they have initial an initial turbulent Mach number ($\calm=5$) that is similar to that in AT10 ($\calm=5.7$). They find that two highly-fragmented disks with Keplerian-like velocity profiles form in their simulation, consistent with our conclusion that normal (i.e., not aligned) turbulence is needed to enable disk formation in a magnetized medium. They suggest that disks can form in turbulent media due to the absence of a coherent rotation structure in the ambient gas and conclude that since misaligned fields are a particular form of disordered fields, their explanation and that of field misalignment \citep{Joos2012,Joos2013} have the same physical origin. They further conclude that the results of their study confirm their previous study \citep{Seifried2012}, which found that the `magnetic braking catastrophe' is a consequence of idealized initial conditions and that turbulence is sufficient to form disks. Finally, \cite{Seifried2015} found that even mild ($M\geq$0.5) turbulence promotes the formation of protostellar disks. They also suggest that Class 0 stage protostars have a greater than 50\% probability of having a disk.

\cite{SantosLima2012,SantosLima2013}, on the other hand, find that the disk formation found in their simulations is consistent with turbulent reconnection. They begin with a uniform medium with a total mass of 1 \msun\ around a stationary 0.5 \msun\ sink particle that also provides the sole gravitational potential. In addition, the gas is initialized with a very fast rotational velocity profile that is $\sim$10 times greater than the observed value \citep{Krasnopolsky2011}. Using an isothermal gas and continuously driven turbulence, the authors find that a large $\sim$75 au disk forms, whereas no disk forms in their rotation-only simulation. By studying the mass-to-flux ratio in the environment around their disk they find that magnetic flux is efficiently diffused away, which they attribute to turbulent reconnection. However, due to the large difference in simulation setup and physics considered---the isothermal equation of state, continuous turbulent driving, high rotational velocity, and lack of self gravity of the gas---it is difficult to make a direct comparison between our results and theirs.

Similar models were performed by \cite{Li2013} using the same model setup as \cite{SantosLima2013}, but varying the initial magnetic field, and thus, the mass-to-flux ratio. They find similar results as \cite{Joos2012}: misalignment is critical to the formation of Keplerian protostellar disks. \cite{Li2013} find that misalignment alone promotes disk formation only in models with $\muphi>$4. However, there is disagreement between \cite{Joos2012} and \cite{Li2013} for a model with a $\muphi$ of 2 and an initial misalignment of 90$^\circ$, where \cite{Joos2012} is able to form a disk while \cite{Li2013} is not. As pointed out by \cite{Li2013}, this discrepancy is likely due to the a difference in the
shape of the magnetic field lines: \cite{Joos2012} adopted a unidirectional field in a centrally concentrated cloud for their initial conditions, whereas \cite{Li2013} adopted a uniform field in a uniform cloud. At the time when \cite{Li2013}'s cloud reached the central concentration of the \citet{Joos2012} cloud, their field had an hour-glass shape that could exert a greater torque than the \cite{Joos2012} field.

Finally, we comment on the resolution used in each set of simulations. As found by \cite{Myers2013}, high resolution is required to develop rotationally supported disks. If the resolution is too low, the disk is found within the accretion region of the sink particles used in these simulations. \cite{Myers2013} found that a 1.25 au resolution is sufficient to resolve these disks. Similar resolution is found in the simulations of \cite{Joos2012} and \cite{Seifried2013}, with 0.4 au and 1.2 au respectively. This also matches well with the resolution used here. The simulations presented in \cite{SantosLima2012}, however, employ a much larger grid resolution of 15.6 au. This does suggest that the resolution used in \cite{SantosLima2012,SantosLima2013} is very coarse compared to other groups and simulations.

Despite the differences between our models and those discussed above, the results of all these studies are broadly consistent: turbulence appears to be essential for the formation of protostellar disks in clouds with mass-to-flux ratios similar to those observed. Here we have shown that if the initial turbulent misalignment is removed, then disks do not form.

\subsection{Comparison with Joos et al. (2013)}

\cite{Joos2013} studied disk formation using a suite of models with a 5 \msun\ molecular cloud and a barotropic equation of state, varying the initial turbulent energy and the mass-to-flux ratio. They used a magnetic field profile that is very similar to ours in that they achieve a constant \muphim\ throughout the cloud and employ a random turbulent velocity field. They find results that are similar to \cite{Seifried2013}, that is, a rotationally supported disk forms with a mass of $\sim 0.1$\msun\ and that disks are easier to form in turbulent environments. For example, for an initial \muphim\ of 5, they find that a massive 1 \msun\ disk forms in their turbulent model whereas a 0.5 \msun\ disk forms in a similar rotation-only model. They attribute this difference to the lowered magnetic braking efficiency in the turbulent simulation.

Since the models of \cite{Joos2013} share much of the same physics as our models, we have run several models with a turbulent, magnetized, 5 \msun\ cloud that aim to reproduce their results. These models are listed in Table.~\ref{tab:simlist} as RT10M5, RT10M5MU5 (which has an initial \muphim\ of 5), AT10M5, and AT10M5L8. \cite{Joos2013} quantify the amount of turbulent energy via $\epsilon = E_{\rm KE}/E_{\rm G}$, the ratio of turbulent kinetic energy to the gravitational energy, which is equal to \alphavirm/2. We concentrate on models with initial \alphavirm$= 0.05-0.1$. Following \cite{Joos2013}, we do not introduce sink particles in this simulation.

RT10M5, RT10M5MU5, and RT10M5L8 employ a random initial turbulence field, \ie\ it has not been aligned with any preferred axis, whereas model AT10M5L8 uses our aligned turbulence pattern. Finally, we use their analytic barotropic equation of state,
\be
P = \rho c_{s,0}^2 \left[ 1.0 + (\rho/\rho_{\rm cut})^{2/3}\right],
\ee
where $\rho_{\rm cut}$ = 1.0$\times$10$^{-13}$ gm cm$^{-3}$ and $c_{s,0}=0.188$ km s$^{-1}$, consistent with an initial temperature of 10 K. The initial density profile takes the form
\be
\rho(r) = \frac{\rho_0}{1.0+(r/r_0)^2},
\ee
where $\rho_0= 1.2\times10^{-18}$ gm cm$^{-3}$ is the central density and $r_0 = 0.023$ pc is the radius of the central region where the density is approximately uniform. The cloud extends to a radius $r_B = 0.07$~pc. These initial conditions are consistent with observations \citep{Andre2000,Belloche2002}.  Since there is no sink particle, the mass of the `star' is estimated as all the gas with density greater than $n_{\rm H}>10^{10}$ cm$^{-3}$, corresponding to $\rho>2.3\times 10^{-14}$ g cm$^{-3}$.

\begin{figure*}
\centering
\includegraphics[trim=0.0mm 0.0mm 0.0mm 0.0mm, clip, scale=0.4]{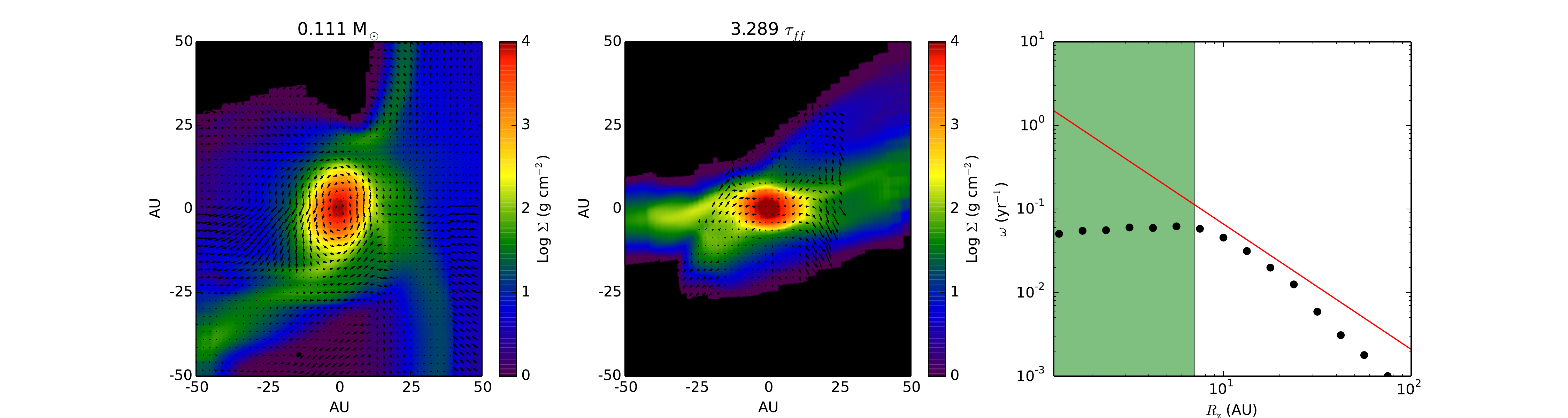}
\includegraphics[trim=0.0mm 0.0mm 0.0mm 0.0mm, clip, scale=0.4]{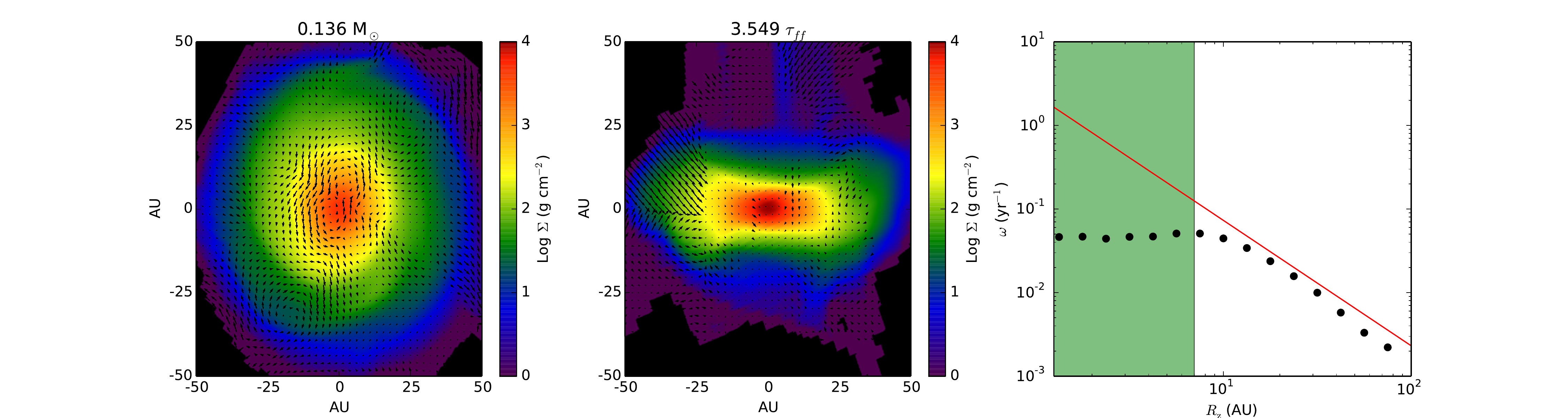}
\includegraphics[trim=0.0mm 0.0mm 0.0mm 0.0mm, clip, scale=0.4]{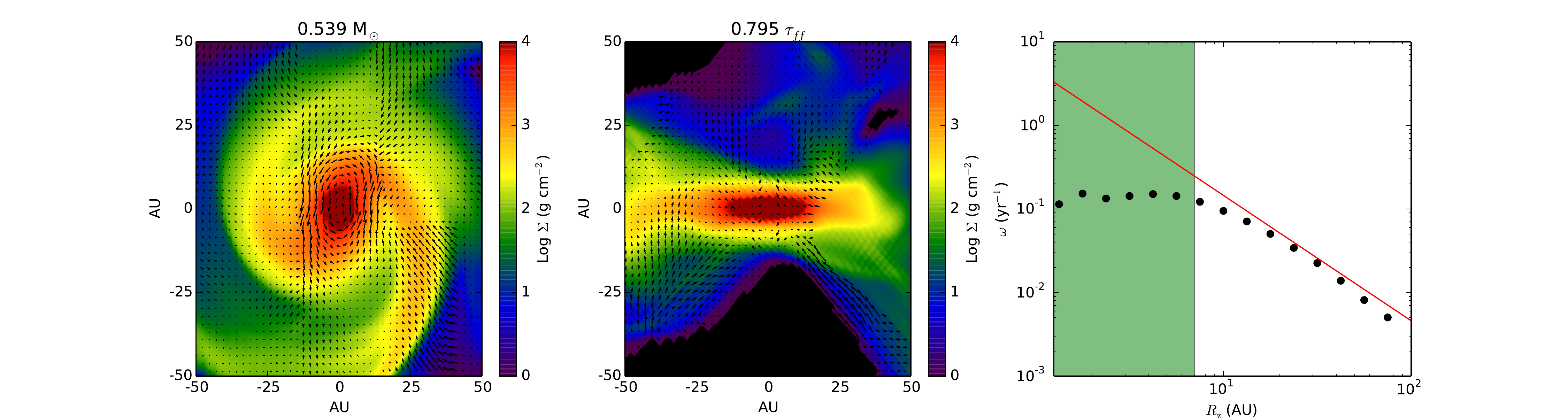}
\caption{ {\it Top Panel}: Density projection of the collapsed region formed in RT10M5, {\it Middle Panel}: AT10M5, and {\it Bottom Panel}: RT10M5MU5. The left column shows the face-on density projection while the middle panel shows the edge-on projections. The right columns show the Keplerian profiles where the black dots show the data values while the (red) line shows the analytic profile. The stellar mass is estimated as all gas with density greater than $\nh>10^{10}$ cm$^{-3}$. The black arrows represent the velocity components.}
\label{fig:small_sc}
\end{figure*}

We find results consistent with those found in \cite{Joos2013}, that is, for RT10M5 and RT10M5MU5, moderate-sized (R$\sim$50 au), rotationally supported disks are formed. Both RT10M5 and RT10M5MU5 meet all the disk criteria out to a radius 50 au; in particular, they each have a very large ratio between the rotational support and thermal pressure (\ie\ $\rho v_{\phi}^2/2P_{\rm thermal}\sim10-50$). Figure~\ref{fig:small_sc} shows the density projections of the collapsed regions formed in RT10M5, RT10M5MU5, AT10M5 with the projected velocity shown as the black arrows. Finally, to ensure our comparison is converged, we reran RT10M5 with an additional level of resolution (RT10M5L8) with the finest level having a minimum size of $\sim$0.8 au. Similar results were found between this run and RT10M5. However, no rotationally supported disk is formed in AT10M5. Specifically, AT10M5 fails to meet both the tangential velocity threshold and, consequently, the rotational support criterion. The presence of disks with random turbulence and the absence of disks with aligned turbulence is consistent with the results found above for the $300\,M_\odot$ clouds.

\subsection{Increase of the Mass-to-flux Ratio Due to Non-ideal MHD Effects}

Several recent studies on the formation of protostellar disks have shown that the mass-to-flux ratio systematically increases with time near the central star as shown in the top row of Figure.~\ref{fig:mvr} \citep[\eg][]{Seifried2013,Joos2013,SantosLima2013} and have suggested several physical mechanisms for the dissipation and loss of magnetic flux.

\citet{Joos2013} suggest that turbulent resistivity is responsible for the diffusion of flux near the disk, which lowers the magnetic braking efficiency and promotes disk formation. The mechanism championed by \cite{SantosLima2013} has been termed ``reconnection diffusion" \citep{Lazarian1999,Lazarian2004,Lazarian2015,Kowal2009,SantosLima2010}. In this scenario, turbulent motions allow for the rapid reconnection of the magnetic field, which increases \muphim\ in centrally concentrated gas.

This is a specific form of turbulent resistivity that is suppressed for sub-\alfvenic\ flows. We also note that for the simulations presented here, the initial turbulence was allowed to decay without any additional driving; this, plus  the transsonic nature of the our turbulence likely means that any possible reconnection diffusion is subdued in our simulations.

 Many of the above studies used simulation setups that either continuously drove the turbulence \cite[\eg][]{SantosLima2010} or had a period of driving before allowing the turbulence to decay \cite[\eg][]{SantosLima2012}.
\cite{Seifried2013} acknowledge that reconnection diffusion could be responsible for the reduction in magnetic flux at $r\la 100$~au seen in the simulations, but they argue that that is insufficient to account for disk formation since (1) the reduction in the flux does not increase \muphim\ above the critical value and (2) magnetic torques must be reduced on scales $\sim 10$ times larger in order for disks to form \citep{Seifried2011}.

\section{Conclusion}

One of the most important phases of star formation is the formation of a rotationally supported gaseous disk. Originally, it was thought that this was a simple consequence of angular momentum conservation during the gravitational collapse of the parent cloud. Observations of molecular clouds, however, have shown that these clouds are relatively highly magnetized, with mass-to-flux ratios $\muphi\simeq 2-3$. The magnetic field exerts a torque on the accreting gas, suppressing disk formation. The ability of the gas to overcome this torque is directly proportional to \muphim\ (\citealp{Mestel1985}; see \S \ref{sec:comp}). For large values of \muphim($\gtrsim 10$) the impact of the field is fairly minor. For magnetic fields as strong as those typically observed, on the other hand, the transfer of angular momentum becomes efficient and the disk is either destroyed or fails to form. This is termed the `magnetic braking catastrophe'. In addition, the inclusion of other non-ideal MHD effects, including the Hall effect, Ohmic dissipation, and ambipolar diffusion, have either a negligible effect or require unphysically large coefficients in order to be effective.

Alternatively, turbulent motions inherent to the cloud have shown promise in overcoming these problems. Other studies, beginning with \cite{Hennebelle2009}, have shown that even modest misalignment between the rotation axis of the molecular cloud and the magnetic field---which turbulence naturally provides---can allow the formation of Keplerian disks. However, the mechanism of disk formation in a turbulent medium, whether a simple consequence of turbulent motions or from `turbulent reconnection diffusion', has remained under active discussion.

We have performed several large-scale MHD simulations of molecular clouds with specialized initial conditions in order to address the turbulence-angle degeneracy as well as study both the `turbulent reconnection diffusion' and turbulent misalignment disk formation mechanisms.
We focused on the case of a realistically strong initial magnetic field, with a mass-to-flux ratio of $\mu_\Phi=2$.
The molecular clouds are initialized with a turbulent velocity field that is initially aligned with the magnetic field.  Using a set of stringent disk criteria \citep{Joos2012}, we find that rotationally supported disks are unable to form when the initial turbulent velocity field is aligned with the magnetic field. We are unable to assess the role of `turbulent reconnection' in the simulations that were the focus of our study since we have sink particles that exclude the magnetic field associated with the gas that forms them. We conclude that turbulence and the associated misalignment between the angular momentum and the magnetic field play a fundamental role in the formation of protostellar disks.

\section*{Acknowledgements}

We would like to thank Andrew Cunningham, Patrick Hennebelle, Shu-Ichiro Inutsuka, Ralf Klessen,  Pak Shing Li, Zhi-Yun Li,  and Andrew Myers  for helpful discussions. We particularly thank Kengo Tomida for pointing out that the angular momentum lost in the growth of sink particles could significantly affect disk formation in numerical simulations and for providing valuable references. Helpful comments by the referee are gratefully acknowledged. This work was performed under the auspices of the U.S. Department of Energy by Lawrence Livermore National Laboratory under Contract DE-AC52-07NA27344 (W.J.G, R.I.K). Support for this work was provided by NASA through ATP grant NNX-13AB84G (R.I.K, C.F.M) and TCAN grant NNX-14AB52G (C.F.M., R.I.K.) and and the NSF through grant AST-1211729 (C.F.M., R.I.K.). Supercomputing support was provided by NASA, through a grant from the ATP, from Lawrence Livermore National Laboratory, and from NERSC. The figures and analysis presented here were created using the {\bf yt} analysis package \citep{Turk2011}.

\appendix

\section{Alignment of the Turbulent Velocity Field}
\label{turbalign}

The inclusion of a turbulent velocity field imparts angular momentum into the cloud \citep{Burkert2000}.
In general the angular momentum vector is pointing in some random direction. One of the goals of this work is to determine whether it is the simple presence of turbulence or the natural misalignment it brings that is important in the formation of disks. To accomplish this, it is important to develop a method that includes turbulence in such a way that the angular momentum vector is aligned closely with the magnetic field. The procedure to do this is described below but can be summarized as follows: First generate a random cube of velocity perturbations of the desired size, then calculate the angular momentum vector and compute the misalignment with the appropriate axis. Next, construct a rotation matrix and rotate both the perturbations and the initial positions using this matrix.  Finally, the final aligned perturbations are interpolated from the rotated perturbations and rotated positions back to the original grid positions.

We generate a velocity perturbation cube with size $N^3$ using the method of \cite{Dubinski1995}. We then define a coordinate system such that the center of the cube corresponds to the origin and each axis spans a range from $-N/2$ and $N/2$. We define a perturbation radius as simply $N/2$. In addition to the cartesian positions $(x,y,z)$, we define a radial position as the spherical radius divided by the perturbation radius. Since the cube will ultimately be placed over a spherical cloud and to facilitate the eventual remap, the perturbation cube is rounded by setting the perturbation values to zero outside the perturbation radius.

The angular momentum vector is then calculated as $\vec{w} = \vec{r}\times m\vec{v}$, where $m$ is the mass at position $r$.  The initial misalignment is then calculated in terms of both an axis and angle as:
\begin{eqnarray}
\vec{R} &=& \vec{A} \times \vec{w}  \nonumber \\
\theta &=& \cos^{-1}( \vec{A}\cdot\vec{w}),
\end{eqnarray}
where $\vec{w}$ is the normalized angular momentum vector and $\vec{A}$ is a unit vector pointing in the desired direction. Here we have defined $\vec{A}$ as $\vec{A} = R_y(\theta)\hat{z}$, where $\hat{z}$ is a unit vector pointing in the positive $z$ direction and $R_y(\theta)$ is the rotation matrix about the y-axis. The user supplies $\theta$. This allows the angular momentum vector to be aligned to any vector that lies in the $x-z$ plane and allows us to specify the amount of misalignment present in the cloud.

Next we construct a rotation matrix using both $\theta$ and $\vec{R}$ and rotate both the initial perturbations and their positions. The final perturbations are then interpolated from the rotated positions and rotated perturbations.

A variety of methods are available to perform the mapping. We have found that for large perturbation cubes (N$>$128), using {$\bf scipy.interpolate.griddata$} with nearest neighbor interpolation gives very accurate results for short compute times. To ensure that the angular momentum is uniformly aligned with a given axis, the rotation-mapping procedure is performed in a series of concentric spherical shells rather than all at once.

\begin{figure}
\centering
\includegraphics[trim=0.0mm 0.0mm 0.0mm 0.0mm, clip, scale=0.40]{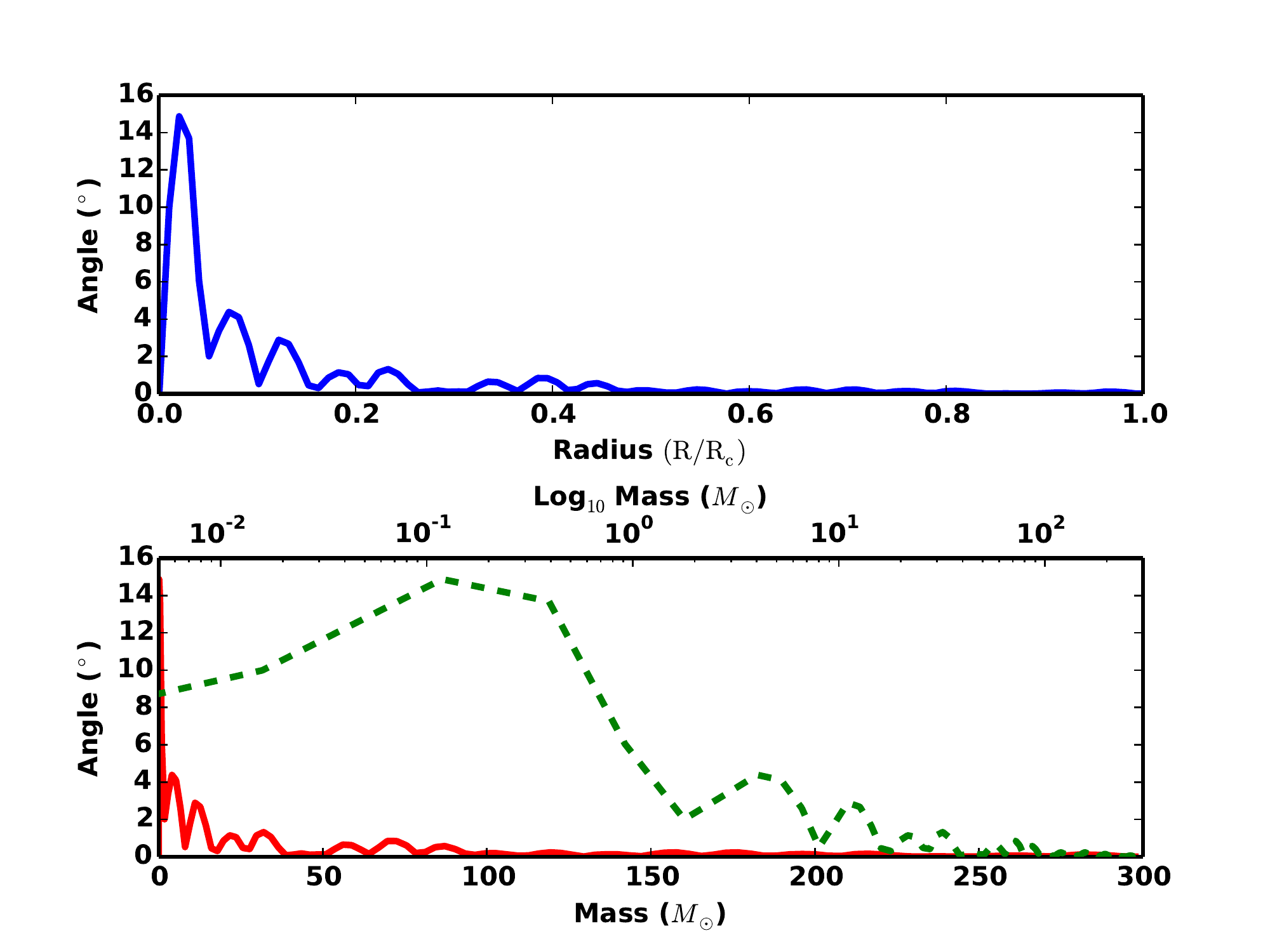}
\caption{Results of the alignment procedure using 20 concentric spherical shells. Misalignment between the $\hat{z}$ and $\vec{J}$ is shown on the y-axes. The top panel shows the misalignment versus the radius of the initial cloud. The bottom panel shows the misalignment versus mass. The bottom axis shows the misalignment versus the linear mass (solid red line) while the top axes shows the misalignment versus the logarithm of the mass (dashed green line). Only at the center does the misalignment exceed 10 degrees and only effects $\sim$ 1 \msun. However, most of the cloud is aligned within 2 degrees of $\hat{z}$.}
\label{fig:jang}
\end{figure}

In the cubes generated here we have used 20 shells to ensure the very center is aligned. Finally, we pad these perturbations with zeros to ensure that the perturbations cover only the simulation domain with the molecular cloud. The results of this procedure are shown in Fig.~\ref{fig:jang}.

\section{Determination of B$_{\rm z}$}
\label{calcmuphi}

The importance of the magnetic field relative to gravity within a cylindrical radius $\varpi$ is expressed by the quantity
$\muphi(\varpi)$, defined as
\be
\mu_{\Phi}(\varpi) = \frac{M(\varpi)}{M_\Phi(\varpi)} = \frac{2\pi \sqrt{G}M(\varpi)} {\Phi(\varpi)}
\label{abmuphi},
\ee
where $G$ is the gravitational constant, $M$ is the mass, and $\Phi$ is the magnetic flux. Our goal is to determine the required magnetic field profile such that for a given density profile, \muphim\ is independent of radius. The required flux at a given radius is given by the mass at a given cylindrical radius,
\be
M(\varpi) = \int_0^{2\pi} d\phi\int_0^{\varpi} \varpi d\varpi\int_{-(R_c^2-\varpi^2)^{1/2}}^{(R_c^2-\varpi^2)^{1/2}} \rho(r) dz
\ee
where
$R_c$ is the cloud radius, $r$ is the radius of a point in spherical coordinates,
and $\rho(r)$ is the spherically symmetric density profile. This profile is numerically integrated to find the mass at a series of radii.
The magnetic field profile is then found from equation (\ref{abmuphi}) under the assumption of a constant \muphim,
\be
B_z(\varpi) = \frac{d\Phi}{2\pi \varpi d\varpi}=\frac{\sqrt{G}}{\mu_{\Phi}} \frac{1}{\varpi} \frac{dM(\varpi)}{d\varpi}   .
\ee
We then fit this profile with a high order polynomial, thereby ensuring that \muphim\ is constant throughout the cloud. In practice we calculate two separate fits, one for the central 5\% of the cloud and one for the remainder. This is done because the density profile is different between these two regimes and, at least initially, the central region of the cloud is highly refined and requires a robust fit.


\bibliographystyle{mnras}
\bibliography{ms}

\bsp	
\label{lastpage}
\end{document}